%% file: ms.tex
\documentclass[12pt,preprint]{aastex}
\begin{document}

\title{The Chemical Composition of Praesepe (M44)}

\author{Ann Merchant Boesgaard\altaffilmark{1}, Brian W. Roper\altaffilmark{1}
\& Michael G. Lum\altaffilmark{1}}

\affil{Institute for Astronomy, University of Hawai`i at M\-anoa, \\ 2680
Woodlawn Drive, Honolulu, HI {\ \ }96822 \\ } 
\email{boes@ifa.hawaii.edu}
\email{brianwroper@gmail.com}
\email{mikelum@ifa.hawaii.edu}

\altaffiltext{1}{Visiting Astronomer, W.~M.~Keck Observatory jointly operated
 by the California Institute of Technology and the University of California.}

\begin{abstract}
Star clusters have long been used to illuminate both stellar evolution and
Galactic evolution.  They also hold clues to the chemical and nucleosynthetic
processes throughout the history of the Galaxy.  We have taken high
signal-to-noise, high-resolution spectra of 11 solar-type stars in the
Praesepe open cluster to determine the chemical abundances of 16 elements: Li,
C, O, Na, Mg, Al, Si, Ca, Sc, Ti, V, Cr, Fe, Ni, Y, and Ba.  We have
determined Fe from Fe I and Fe II lines and find [Fe/H] = +0.12 $\pm$0.04.  We
find that Li decreases with temperature due to increasing Li depletion in
cooler stars; it matches the Li-temperature pattern found in the Hyades.  The
[C/Fe] and [O/Fe] abundances are below solar and lower than the field star
samples due to the younger age of Praesepe (0.7 Gyr) than the field stars.
The alpha-elements, Mg, Si, Ca, and Ti, have solar ratios with respect to Fe,
and are also lower than the field star samples.  The Fe-peak elements, Cr and
Ni, track Fe and have solar values.  The neutron capture element [Y/Fe] is
found to be solar, but [Ba/Fe] is enhanced relative to solar and to the field
stars.  Three Praesepe giants were studied by Carrera and Pancino; they are
apparently enhanced in Na, Mg, and Ba relative to the Praesepe dwarfs.
The Na enhancement may indicate proton-capture nucleosynthesis in the Ne
$\rightarrow$ Na cycling with dredge-up into the atmospheres of the red
giants.
\end{abstract}

\keywords{stars: abundances; stars: evolution; stars: late-type; stars:
solar-type; open clusters and associations: general; open clusters and
associations: individual (M 44, NGC 2632, Praesepe); Galaxy: evolution}

\section{INTRODUCTION}

Open clusters have provided important information in the study of stellar
evolution, Galactic chemical evolution, nucleosynthesis and light element
abundances.   The stars in a given cluster are formed from the pre-cluster gas
with similar compositions and ages.

The Praesepe cluster is about the same age as the Hyades cluster at 0.7 Gyr
(e.g.~Salaris et al.~2004).  Metallicity determinations cover a range of
values.  Boesgaard (1989) determined [Fe/H] for several open clusters
using sharp-lined cluster stars with the best high-resolution spectra for each
cluster; for Praesepe she used the Palomar 5-m telescope and derived [Fe/H] =
+0.09 $\pm$0.07.  Friel \& Boesgaard (1992) found [Fe/H] = +0.04 $\pm$0.04
from six sharp-lined F dwarfs observed with CFHT high-resolution, high
signal-to-noise spectra.  With high-resolution spectra of four G dwarfs in
Praesepe An et al.~(2007) derived [Fe/H] = +0.11 $\pm$0.03.  With spectra from
VLT + UVES Pace et al.~(2008) found abundances for 6 elements in 7 Praesepe
dwarf stars; their Fe abundances are supersolar at +0.27 $\pm$0.10.  The
compilation of Gratton (2000) gives +0.04 $\pm$0.06 and Salaris et al.~(2004)
use +0.13 $\pm$0.06 for Praesepe as one of their calibrating clusters.
Carrera \& Pancino (2009) found [Fe/H] = +0.16 $\pm$0.05 for three red giant
stars in Praesepe.

The Hyades and Praesepe clusters have long been thought to be so similar as to
be coeval having formed in the same giant molecular cloud complex.  In their
study of the places of origin of 24 open clusters, Palous et al.~(1977) find
that for the angular rotational speeds of 13.5, 15.0, 17.5 and 20.0 km
s$^{-1}$ kpc$^{-1}$ Hyades and Praesepe were formed near each other.  Their
metallicities are similar at [Fe/H] $\sim$+0.13, their ages are similar at 0.7
Gyr (e.g. Salaris et al.~2004, Magrini et al.~2009) and their kinematic
properties are similar (Eggen 1992).

There are some intriguing differences.  The activity level in Praesepe is
lower than that in the Hyades.  ROSAT studies of the Hyades showed a detection
rate of 90\% for the G dwarfs in the Hyades (Stern et al.~1995), but only 33\%
in the Praesepe G dwarfs (Randich \& Schmidt 1995).  The dichotomy in the
X-ray luminosity functions is not due to membership problems, sensitivity
issues or differences in rotational velocity distributions (Barrado y
Navascues et al.~1998).  Holland et al.~(2000) and Franciosini et al.~(2003)
suggest that Praesepe may actually be two merging clusters of different ages.
The former authors describe the main cluster with 630 M$_{\odot}$ extending to
12.1 pc with a subcluster of 30 M$_{\odot}$ that is 3 pc away from the center.
There might be a possibility of somewhat different Fe abundances in the
two merging pieces, but the major part is dominant by a factor of 21 in mass.

In this paper we present the abundances of 16 elments in 11 solar-temperature
dwarfs in Praesepe from high-resolution, high signal-to-noise spectra obtained
at the Keck I telescope with HIRES.  We compare our results with those of
Carrera \& Pancino (2011) who have determined chemical abundances in three red
giants in Praesepe.

\section{OBSERVATIONS AND DATA REDUCTION}

The stars selected for this work are main sequence stars with colors and
temperatures surrounding the solar value of 5774 K.  High-resolution spectra
were obtained with HIRES (Vogt et al.~1994) on the Keck I telescope on Mauna
Kea on two clear nights in January and February 2003.  The spectral resolution
is $\sim$45,000 and the signal-to-noise ratios (S/N) range from 93 to 166 with
a median of 153.  Typical integration times were 20 - 25 minutes.  The details
of the observations of 11 Praesepe stars are given in Table 1.  The spectral
coverage is from 5730 \AA\ to 8140 \AA\ with some interorder gaps.  On the
night of 2003 January 11 (UT) we also took a 10 s exposure of the Moon to be
used as a surrogate for the solar spectrum.

Each night we obtained 13 flatfield frames and 13 bias frames as well as Th-Ar
spectra at the beginning and end of the night for wavelength calibration.  The
data reduction was done using standard IRAF\footnote{IRAF is distributed by
the National Optical Astronomical Observatories, which are operated by AURA,
Inc.~under contract to the NSF.} routines.  These include
overscan-subtraction, bias-subtraction, master nightly flat-field
normalizations, wavelength calibrations, scattered-light removal, cosmic ray
removal, and continuum fitting.  We were able to extract 19 orders of spectra.

In Figure 1 we show the color-magnitude diagram for Praesepe from UBV
photometry done by Johnson (1952) and Mendoza (1967).  The stars we observed
in Table 1 are indicated by the open circles.  All of our stars are
confirmed members based on radial velocity measurements of Mermilliod \& Mayor
(1999).  All the stars are confirmed members (at 97-99\%) from proper motion
measurements of Jones \& Cudworth (1983) and Jones \& Stauffer (1991), except
KW 30 which they did not measure.

Figure 2 shows 60 \AA\ of spectrum for three of our stars which span a range
in temperature.  Several Fe I lines used in the analysis are indicated.  The
high quality of our data can be seen in this figure.  It is easy to see that
as the temperature decreases, the Fe I lines strength increases.  Figure 3
shows the region near the Li I line for the same three stars.  The Li I line
decreases in strength going to cooler temperatures due to increasing depletion
of Li.  The three Fe I lines in that figure increase in strength with
decreasing temperature.  Figure 4 shows the region where the high excitation C
I lines occur while Figure 5 is of the high excitation lines of the O I
triplet in the same three stars.

\section{ABUNDANCES}

We have used both IRAF and
MOOG\footnote{http://www.as.utexas.edu/~chris/moog.html} (Sneden 1973, as
revised in 2002) to analyze the reduced spectra.  Equivalent widths were
measured with the {\it splot} task in IRAF for each star.  The line list
we used is given in the Appendix.  We edited the Fe I and Fe II line lists to
omit lines weaker than 5 m\AA{} that might have poorer quality measurements.
Most of the line lists for Fe and the other elements are from Stephens
(1999), Stephens \& Boesgaard (2002), Reddy et al.~(2003), Kurucz (1995) and
the NIST data site.  Additions to those lists are described in the relevant
parts of $\S$4.

\subsection{Stellar Parameter Determination}

We have used the infra-red flux method (IRFM) to find $T_{\rm eff}$.  Table 2
gives the measured colors for our stars, primarily from 2MASS.  Table 3 gives
the temperatures derived from the color indices and the calibration of
Casagrande et al.~(2010).

With the exception of Sc II and Ba II, our abundances are not very sensitive
to the value of log g.  Inasmuch as we are dealing with stars in a cluster, we
have used the relationship between log g and B-V compiled by Gray (1976) from
45 main-sequence eclipsing-binary stars.  We have used the empirical
relationship for microturbulent velocity derived by Edvardsson et al.~(1993)
with its dependency on log g and $T_{\rm eff}$.

We determined values for [Fe/H] from 50-55 lines of Fe I and 5-6 lines of Fe
II.  We found the mean [Fe/H] by weighing Fe I and the Fe II results by the
number of lines measured for each.  In Table 4 we show these results for
[FeI/H] and [FeII/H] along with the average deviation from the individual
lines for each star.  The final $<$[Fe/H]$>$ was +0.117 $\pm$0.039.  In the
subsequent models for all the stars we used [Fe/H] = 0.12.  Table 5 gives the
adopted model parameters for the 11 stars in this study.

\subsection{Abundance Determinations}

We used Kurucz (1993) model atmospheres and interpolated among his grid models
to create a model atmosphere for each star.  For Li we used the {\it synth}
driver in MOOG to make synthetic spectra to determine the Li abundance.  For
the other elements we used measurements of equivalent widths and the {\it
abfind} driver in MOOG, with the exception of Ba II for which we include the
hyperfine splitting (hfs) and use the {\it blends} driver.  (The Appendix
contains a table giving the wavelengths, excitation potentials, log gf values
and measured equivalent widths of the lines used for the three stars shown in
Figures 2 -- 5.)  The final abundances are given in Table 6.

\subsection{Abundance Uncertainties}

We estimated errors due to uncertainties in the stellar parameters.  Table 3
shows the agreement among the temperatures determined from the various color
indices.  The uncertainty in the mean ranges from 29 to 111 K with a mean of
56 K, median of 50 K.  We chose $\pm$75 K as a conservative estimate of the
uncertainty in $T_{\rm eff}$.  The average deviation from the linear relation
from Gray between log g and B-V is $\pm$0.09 dex.  Our choice of $\pm$0.20 dex
for the uncertainty in log g is also conservative.  Our mean [Fe/H] of
$\pm$0.117 has a standard deviation of $\pm$0.039, so our use of $\pm$0.10 in
[Fe/H] is again conservative.  The Edvardsson et al.~(1993) relation for
microturbulent velocity has an rms scatter of 0.3 km s$^{-1}$ based on 157
field stars; for cluster stars of virtually the same log g (4.38 - 4.43) the
rms scatter is considerably less and we used $\pm$0.20 km s$^{-1}$.  Table 7
shows those abundance errors for Fe I.  We added the errors in quadrature to
estimate the total error due to the parameter uncertainties.  In Table 8 we
show the errors for the three representative stars shown in Figures 2-5 for
the other elements.

\section{RESULTS}

\subsection{Iron}

The values for [Fe/H] were given in Table 4.  These are plotted in Figure 6
where the errorbars shown are due to the parameter uncertainties for each star
from Table 7.  We have derived the values of [Fe/H] = +0.117 $\pm$0.039 (the
sample standard deviation).  Including the fact that there are 11 stars in our
sample, the error is $\pm$0.012.

Our value for [Fe/H] of +0.12 is in good agreement with that of An et
al.~(2007) who found [Fe/H] = +0.11 $\pm$0.03 from four G dwarfs; there are
two stars in common with this study, KW 23 and KW 58, which agree better than
within the quoted errors.  We do not agree with Pace et al.~(2008) whose
[Fe/H] abundance from seven stars was supersolar at +0.27 $\pm$0.10.  Salaris
et al.~(2004) used +0.13 for [Fe/H] and Carrera \& Pancino (2011) found +0.16
$\pm$0.05 from three giants stars in Praesepe.

\subsection{Lithium}

We have used MOOG with the {\it synth} driver to find Li abundances, A(Li) =
log N(Li) +12.00.  The line list used for the Li synthesis is from King \&
Hiltgen (1996).  Figure 7 shows the abundance matches for two of our stars.
For all 11 stars the syntheses were excellent fits to the data.  Lithium
abundances or upper limits have been determined by Soderblom et al.~(1993) for
63 Praesepe stars with temperatures between 4970 and 6810 K from Li equivalent
width measurements; these have been redetermined by them in Soderblom et
al.~(1995).  Boesgaard \& Budge (1988) found Li abundances in seven additional
Praesepe F dwarfs.  In 1995 Balachandran recalibrated the temperature scale
and excluded photometric binaries and double-lined spectroscopic binaries from
those samples leaving 59 stars.  Our synthesized spectral results agree well
with those that Balachandran (1995) reexamined via equivalent width
measurements.  The mean difference in A(Li) is +0.07 dex and those differences
are result from the temperature differences.  Similarly, the agreement with
the Soderblom et al.~revised Li abundances is good with the mean difference in
A(Li) = +0.04.  We have six stars in common with King \& Hiltgen (1996) where
our mean A(Li) difference is +0.05.

Soderblom et al.~(1993, 1995) found that the low mass stars in the Praesepe
cluster had higher Li abundances than the low mass stars in the Hyades with
T$_{\rm eff}$ $<$ 5800 K.  Once Balachandran (1995) put the two clusters on
the same temperature scale, that Li abundance difference disappeared.  We show
in our Figure 8 that Hyades and Praesepe have virtually the same Li abundances
in the temperature range of our Praesepe observations: 5650 - 6000 K.  Our
Hyades Li abundances are taken from Boesgaard et al.~(in preparation)
re-evaluation of the Hyades Li abundances with the stellar parameters
determined from the new Hipparcos calibration.

\subsection{Carbon and Oxygen}

Both C and O are formed from massive stars ($\gtrsim$10M$_{\odot}$) and --
through core-collapse supernovae -- have enriched the gas out of which the
early generations of stars were formed. These, in turn, have added their own
contributions to successive generations.  The bulk of the Fe is formed in
intermediate mass stars in supernovae type Ia, so the production of Fe lags
behind that of C and O, e.g.~Tinsley (1980), Wheeler et al.~(1989).  This
leads to the expectation that [C/Fe] and [O/Fe] will be greater than zero in
metal-poor stars.  A recent study that included both Fe and O in 117 stars
with [Fe/H] from $-$0.5 to $-$3.5 by Boesgaard et al.~(2011) shows a monotonic
decrease in [O/Fe] with [Fe/H].  At [Fe/H] = $-$3.5 the value for [O/Fe] is
+1.0 and by [Fe/H] = $-$0.5 the value for [O/Fe] has declined to +0.2.  A
similar trend was found by Boesgaard et al.~(1999) and Israelian et al.~(1998,
2001).

The seven lines of C I and three lines of O I (the O triplet) are all lines
with high excitation potential.  The gf values for C I are from Weise et
al.~(1996) and Reddy et al.~(2003) while the O I triplet line gf values are
from Weise et al.~(1996).  The C and O abundances in our Praesepe dwarfs are
given in Table 6.  The O abundances given there have been corrected for NLTE
effects through the calculations of Takeda (2003).  Rentzsch-Holm (1996) has
studied the NLTE abundance corrections for C in stars with temperatures 7000
-- 12,000 K, log g values of 3.5, 4.0, 4.5 and metallicities of [M/H] of
$-$0.5, 0.0, +0.5, and +1.0.  These trends indicate that for our
solar-temperature stars with C I equivalent widths of 10 -- 32 m\AA, the NLTE
effects are insignificant.  We can compare our C and O abundances with those
in field stars.  We have selected the field stars from Edvardsson et
al.~(1993) for O comparisons and those in Takeda \& Honda (2005) for C and O.
In both comparison samples we have restricted the range in [Fe/H] to be 0.00
to +0.20, enveloping our Praesepe range of +0.05 to +0.17.  In addition we
have used the results from Reddy et al.~(2003, 2006) for C in F and G dwarfs
in the thin disk and the thin-thick disk in our metallicity and temperature
range.  None of the comparison stars is as young as Praesepe.

Figure 9 shows the ratio [C/Fe] for the Praesepe stars in the top panel along
with the Reddy et al.~(2003, 2006) and Takeda \& Honda (2005) comparisons.
By the solar age the excess of C over Fe is expected to have disappeared.  In
the young Praesepe cluster (age $\sim$0.7 Gyr) the value we find for [C/Fe] of
$-$0.14 $\pm$0.07 could indicate the continued decrease in [C/Fe] with time.

The lower panel of Figure 9 shows [O/Fe] as corrected for NLTE effects for
Praesepe and the two comparison samples of Edvardsson et al.~(1993) and Takeda
\& Honda (2005).  (Edvardsson et al.~(1993) did not do a correction for NLTE
effects, but rather calibrated their abundances found from the O I triplet
alone to the abundances found in those stars where they had results from both
the [O I] line at 6300 \AA{} and the O I triplet.)  As is the case for [C/Fe],
we find that O has decreased relative to Fe and the mean value of [O/Fe] is
$-$0.15 $\pm$0.09.  The cluster stars have O abundances typically below the
field star sample.

\subsection{Alpha-Elements: Mg, Si, Ca, Ti}

For the abundance determinations we have used two lines of Mg I, 11 lines of
Si I, 10 lines of Ca I, and 11 lines of Ti I as listed in the Appendix.  The
alpha-element abundances are given for each star in Table 6.  Their ratios
normalized to Fe are shown in Figures 10 and 11 along with the mean values for
[X/Fe] and [Fe/H] for Praesepe.  The comparison samples are from Edvardsson et
al.~(1993) and Reddy et al.~(2003, 2006).  The field stars, except for Ca,
show a larger range in abundance than do the Praesepe stars.  And the Praesepe
stars, again except for Ca, have lower values of [X/Fe] than the field stars.
The alpha-element abundances relative to Fe (relative to solar) tend to
decrease with age reaching solar at the age of the Sun.  The Praesepe stars
are all younger than the field star sample and do show solar abundances.  The
alpha ratios relative to Fe are all close to solar: [Mg/Fe] = $-$0.003
$\pm$0.040; [Ti/Fe] = $-$0.035 $\pm$0.062; [Ca/Fe] = $-$0.006 $\pm$0.049;
[Si/Fe] = $-$0.005 $\pm$0.031.  The mean ratio, [$<\alpha>$/Fe], is $-$0.012
$\pm$0.015.

According to Tsujimoto et al.~(1995) the contribution of SNe Ia to the solar
abundances is only $\sim$1\% for Mg, $\sim$25\% for Ca, and $\sim$17\% for Si.
With the exception of Mg, both SNe Ia and SN II have contributed to the
alpha-elements in the Praesepe cluster.

\subsection{Fe-Peak Elements: Cr and Ni}

In addition to the 55 lines of Fe I and six lines of Fe II, we have measured
nine lines of Cr I and 26 lines of Ni I.  Abundances for Cr and Ni are given in
Table 6 for each star along with the mean abundances relative to Fe and the
standard deviation of the mean.  Figure 12 shows the results as a function of
[Fe/H] with the comparison samples of field stars.  Bergemann \& Cescutti
(2010) studied the effects of NLTE on Cr I in the Sun and metal-poor stars.
Our Cr abundances have been corrected for the overionization effect on Cr I
using the solar value of log $\epsilon$(Cr II)$_{\odot}$ = 5.77 from Sobeck et
al.~(2007).  The Praesepe cluster mean, [Cr/Fe] = 0.003 $\pm$0.031, is the
solar value and in agreement with the field star sample.  According to Clayton
(2003) both SN II and SN Ia produce Cr/Fe ratios that are roughly solar.

The mean value we find for [Ni/Fe] is also similar to the solar value at
$-$0.028 $\pm$0.027.  The abundances are comparable to the field stars from
Reddy et al.~(2003, 2006), but seem lower than the bulk of the Edvardsson et
al.~(1993) field stars of comparable metallicity.  Our two iron-peak elements
behave as Fe does.

\subsection{n-Capture Elements: Y and Ba}

We have determined abundances for the two n-capture elements, Y and Ba, which
are dominated by the s-process at different s-process peaks.  For Y we have
measured two lines of Y I and three lines of Y II, but not all lines were
measurable in most of the stars; we also measured these lines in our lunar
spectrum.  The gf values for Y II are from Hannaford et al.~(1982).
We then normalized our stellar Y abundance to our lunar/solar Y to get [Y/H]
for the Praesepe stars.  For Ba we had one line of Ba II at 5853.7 \AA{} for
which we included the hyperfine structure in our analysis.  The results for
[Y/Fe] and [Ba/Fe] are given in Table 6.  Figure 13 shows the results for
[Y/Fe] (upper panel) and [Ba/Fe] (lower panel) with the field star comparison
samples.  The field stars show a large range ($\pm$0.20) in both elements; our
[Y/Fe] vales are solar near the middle of the comparison stars.  Our [Ba/Fe]
are somewhat higher than solar at +0.11 $\pm$0.04 and near the top of the
field star results.  This is consistent with the results of D'Orazi et
al.~(2009) who found that [Ba/Fe] increases in open clusters with younger
ages.  They suggest that the enhancement of [Ba/Fe] would come from low mass
stars ($\lesssim$1.5M$_{\odot}$).

\subsection{Other Elements: Na, Al, Sc, V}

We also found abundances for some other elements of interest.  We have
measured equivalent widths of two lines of Na I, one Al I line, two lines of
Sc II, and two lines of V I.  (Reddy et al.~(2003) showed in test calculations
that the effects of hyperfine splitting on the lines they used of V I and Sc
II have virtually no effect on the abundances derived; we used those same
lines.)  Figure 14 shows the results for Na and Al.  The value for [Na/Fe] and
[Al/Fe] are solar at $-$0.011 $\pm$0.032 and 0.004 $\pm$0.038, respectively,
and are in good agreement with the field star sample.  See $\S$5 for an
interesting comparison with Na in the Praesepe giants.

In Figure 15 we show the results for V and Sc.  The ratios of [V/Fe] and
[Sc/Fe] are somewhat greater than solar at +0.035 $\pm$0.053 and 0.036
$\pm$0.024, respectively, but they are basically solar within the errors.  The
field stars from the Reddy papers are similar, but, as expected, they have
greater scatter compared to the cluster stars.

\section{COMPARISON WITH PRAESEPE GIANTS}

Carrera \& Pancino (2011) determined abundances of many elements in three red
giants in Praesepe.  We can compare our results for solar-temperature dwarf
stars with theirs for the red giants.  Table 9 and Figure 17 show those
comparisons.  Our values for [Fe/H] are in good agreement within our 1 sigma
errors.  

Their results for [Na/Fe] show an enhancement to +0.25 $\pm$0.06 from our
solar value ([Na/Fe] = $-$0.01 $\pm$0.03).  Our two studies have used the
same Na lines and the same gf values.  The lines selected are weak Na lines
and expected to have only minor NLTE corrections, e.g. Asplund (2005).  Each
of their three red giants shows the enhancement of [Na/Fe]: +0.23, +0.30,
+0.18.  Such a Na-enhancement could be the result of proton-capture
nucleosynthesis in the Ne $\rightarrow$ Na cycling with dredge-up into the
atmospheres of the red giants.  Enhanced Na and an anti-correlation between Na
and O has been discovered recently in evolved red giants in the populous old
open cluster, NGC 6791, by Geisler et al.~(2012).  There are too few
stars in Praesepe that have evolved to become red giants to detect any such
effect in Praesepe.  All three giants studied by Carrera \& Pancino (2011) do
have low [O/Fe] at $-$0.11 $\pm$0.03.

For the alpha elements, Si, Ca, and Ti, the agreement is very good and close
to solar in both dwarfs and giants.  However, Carrera \& Pancino (2011)
found [Mg/Fe] is enhanced to +0.27 in the giants.  Our value for the dwarf
stars is solar.  Their three giants give similar values for [Mg/Fe]: +0.22,
+0.27, +0.31.  Our two studies have measured different lines of Mg I.  Two of
their four lines are quite strong at more than 100 m\AA.  It is not expected
that Mg would be increased in the giants, rather the Mg $\rightarrow$ Al
cycling would result in a decrease in Mg.  Yong et al.~(2003) find a positive
correlation between Al and Mg for some giant stars in the globular cluster,
NGC 6752, apparently due to an increase in $^{26}$Mg (see their Figure 12).
It might be interesting to determine the Mg isotope ratios in those giant
stars.  On the other hand, the [Al/Fe] abundances in both the dwarfs and the
giants in Praesepe are solar.

Our Ba abundances are from the Ba II line at 5853 \AA{} in which we included
the hyperfine structure of that line.  Our mean cluster abundance for
[Ba/Fe] is +0.11 $\pm$0.04 for the main sequence stars which is consistent
with the findings of D'Orazi et al.~(2009) discussed in $\S$4.6.  Carrera \&
Pancino (2011) found [Ba/Fe] is +0.33 $\pm$0.05 for their three giants.  This
could be evidence for n-capture-enriched Ba in the giants.  However,
apparently Carrera \& Pancino (2011) did not use the hyperfine splitting (hfs)
in their Ba analysis.  In order to determine the effect of the hfs on the Ba
abundance, we determined Ba abundances without including the hfs for our 11
dwarf stars.  We find that including the hfs reduces the Ba abundance by 0.03
dex.  If the correction is similar for giants, then [Ba/Fe] would be +0.30 in
those three giants; this is still an increase over [Ba/Fe] in the dwarfs.

\section{SUMMARY AND CONCLUSIONS}

We have analyzed high-resolution, high S/N spectra of 11 solar-temperature
main sequence stars in the Praesepe open cluster obtained with Keck I + HIRES.
The spectra cover a region from 5730 -- 8140 \AA.  We have determined stellar
temperatures from the infrared flux method and found Fe abundances from some
55 Fe I lines and 6 Fe II lines in each star.  The [Fe/H] agreement between
the Fe I and Fe II lines is excellent and the cluster mean is [Fe/H] = 0.12
$\pm$0.04 ($\pm$0.01).

We determined Li abundances by the spectral synthesis method and found them to
track the Hyades Li abundances very well showing a steady decline from 6000 K
to 5650 K; the decline is due to increasing Li depletion at decreasing
temperatures.  We find that [C/Fe] and [O/Fe] to be $-$0.14 $\pm$0.07 and
$-$0.15 $\pm$0.09, respectively.  These values are similar, but somewhat lower
than the field star samples and we interpret the lower values to be due to the
younger age of Praesepe relative to the field stars.  This follows the steady
decline in [C/Fe] and [O/Fe] over time from the early excess of C and O over
Fe as produced by the most massive stars and SN II followed by the rise in Fe
from SN Ia.  All the $\alpha$-elements, [Mg/Fe], [Si/Fe], [Ca/Fe], and [Ti/Fe]
are solar at 0.00 $\pm$0.04, $-$0.01 $\pm$0.03, $-$0.01 $\pm$0.05, and $-$0.04
$\pm$0.06, respectively.  As is the case for [C/Fe] and [O/Fe], the
$\alpha$-elements, Mg, Ti, and Si, are somewhat lower relative to Fe compared
to the field stars.  For [Ca/Fe] the field star sample is also solar and has
less spread in the values.

The abundances of the Fe-peak elements, [Cr/Fe] and [Ni/Fe], were found to
track Fe and are basically solar at 0.00 $\pm$0.03 and $-$0.03 $\pm$0.03,
respectively.  The n-capture element Y from the first s-process peak was found
to be solar at [Y/Fe] = 0.01 $\pm$0.08 and in good agreement with the field
stars.  The n-capture element Ba from the second s-process peak was found to
be enhanced relative to solar and relative to the field stars with [Ba/Fe] =
+0.11 $\pm$0.04.  For the Praesepe stars [V/Fe], [Sc/Fe], [Na/Fe], and [Al/Fe]
are essentially solar at 0.04 $\pm$0.05, 0.04 $\pm$0.02, $-$0.01
$\pm$0.03, and 0.00 $\pm$0.04, respectively. Both [Na/Fe] and [Al/Fe] are
somewhat lower than the field stars and have less spread in the values, as
expected in a cluster of stars of common origin.

The comparison with the composition of the giant stars in Praesepe yielded
some interesting differences for Na, Mg, and Ba.  The three red giants studied
by Carrera \& Pancino (2011) show an enhancement in [Na/Fe] of +0.26 compared
to our dwarf stars.  The enhancement of Na might be caused by the
proton-capture nucleosynthesis in the Ne $\rightarrow$ Na cycling in the
interior and subsequent dredge-up in the giant stars.  For the
$\alpha$-elements Si, Ca, and Ti the dwarfs and the giants are similar within
the errors.  However, they found an enhancement in [Mg/Fe] in the giants of
+0.27.  If there is Mg $\rightarrow$ Al cycling, that would lead to a decrease
in [Mg/Fe] and an increase in [Al/Fe]; for both dwarfs and giants [Al/Fe] =
0.00 with similar uncertainties $\sim$$\pm$0.04.  It may be important to
measure the Mg isotopes in the giants in case the increase in Mg is due to an
increase in $^{26}$Mg.  Barium appears to be enhanced in the giants as well.
We find [Ba/Fe] = +0.11 $\pm$0.04 for the dwarf stars while they derive +0.33
$\pm$0.05.  The giants may be enriched by the s-process n-capture.  However,
there is no apparent enrichment of [Y/Fe] in the giants.

\acknowledgements We would like to thank John Lakatos for his help with the
observing and data reduction and Hai Fu for his IRAF routine to expedite
equivalent width measurements.  We acknowledge support from NSF through grant
AST 05-05899 to A.M.B.

\clearpage
\input{tab1.tex}


\clearpage
\input{tab2.tex}

\clearpage
\input{tab3.tex}

\clearpage
\input{tab4.tex}

\clearpage
\input{tab5.tex}

\clearpage
\input{tab6.tex}

\clearpage
\input{tab7.tex}

\clearpage
\input{tab8.tex}

\clearpage
\input{tab9.tex}

\clearpage
\begin{figure}
\plotone{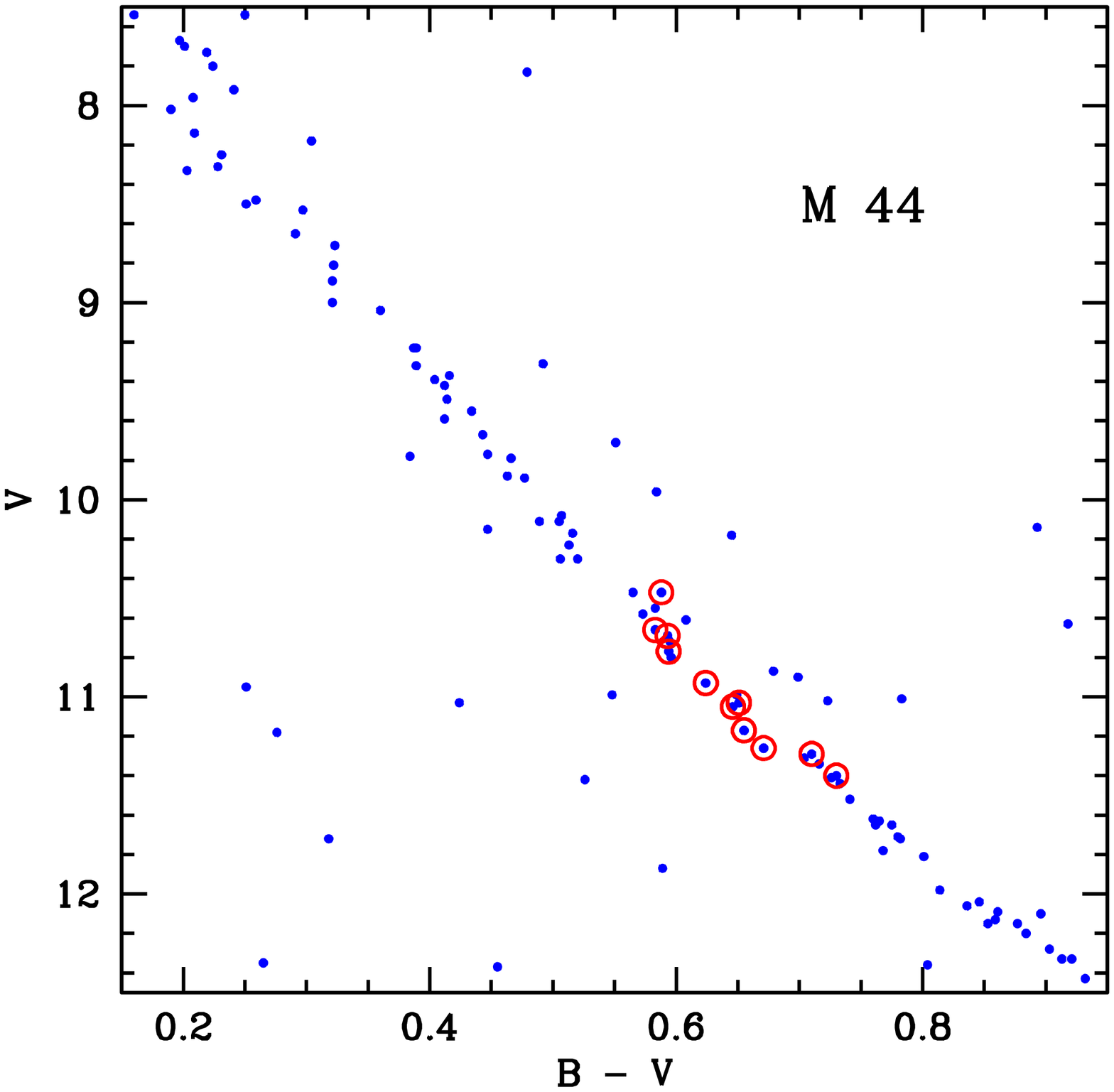}
\caption{The color-magnitude diagram for Praesepe.  The data are from Johnson
(1952) and Mendoza (1967).  The stars we have observed are circled.}
\end{figure} 

\begin{figure}
\plotone{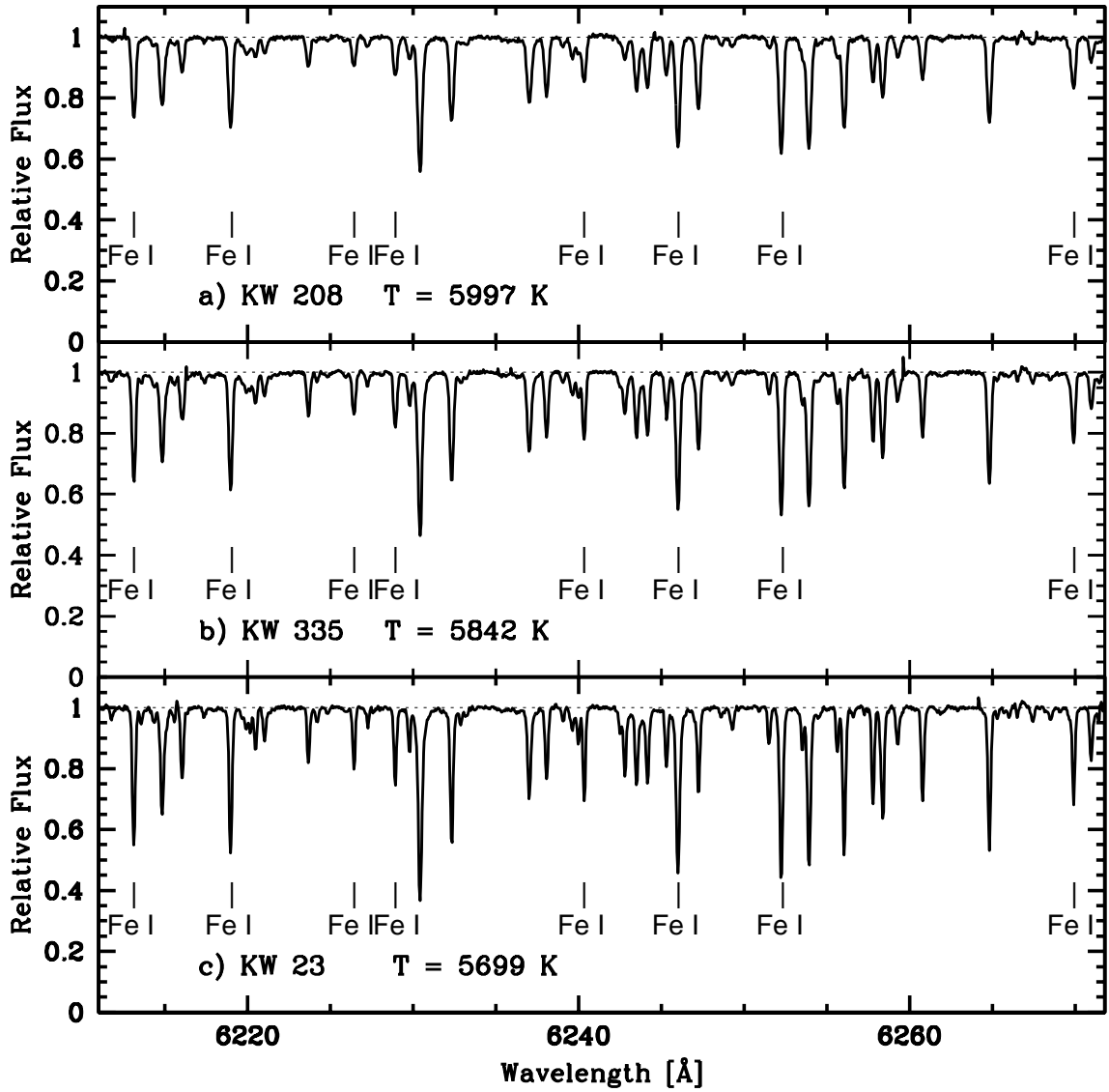}
\caption{Samples of our spectra in the 6200 \AA\ region.  The three stars
cover a range in temperature and the Fe I lines indicated clearly become
stronger as the temperature decreases.  The S/N per pixel for these spectra is
$\sim$150.  The continuum is shown as a light dotted line at 1.0.}
\end{figure}

\begin{figure}
\plotone{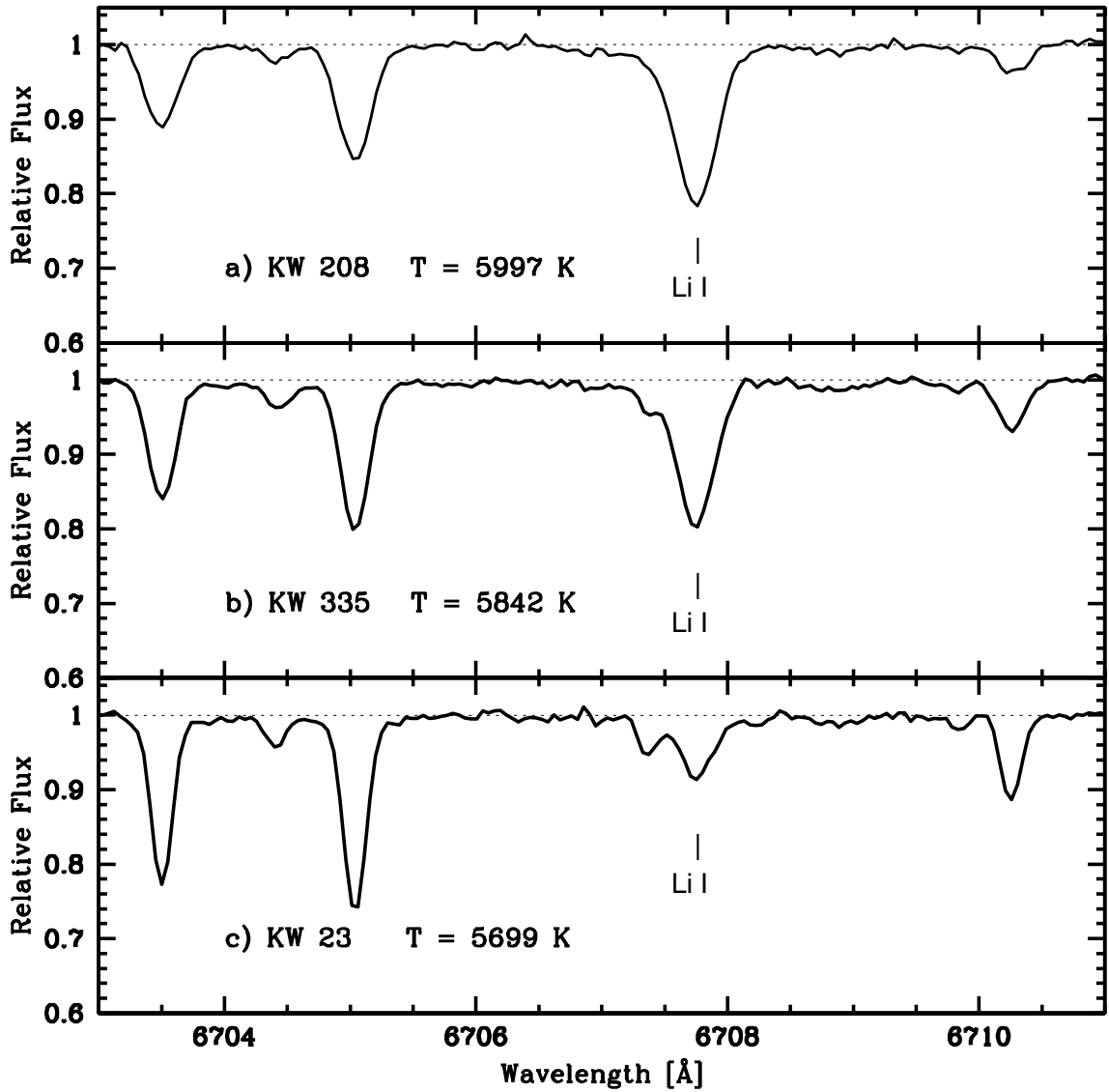}
\caption{Samples of our spectra in an 8 \AA\ region around the Li I doublet.
With decreasing temperature the Li I feature decreases due to Li depletion
while the blending line increases in strength.  The lines at 6703, 6705, and
6710 \AA\ are due to Fe I.  The continuum is shown as a light dotted line at
1.0.  The vertical scale in this figure covers a smaller range than the one in
Figure 2.}
\end{figure}

\begin{figure}
\plotone{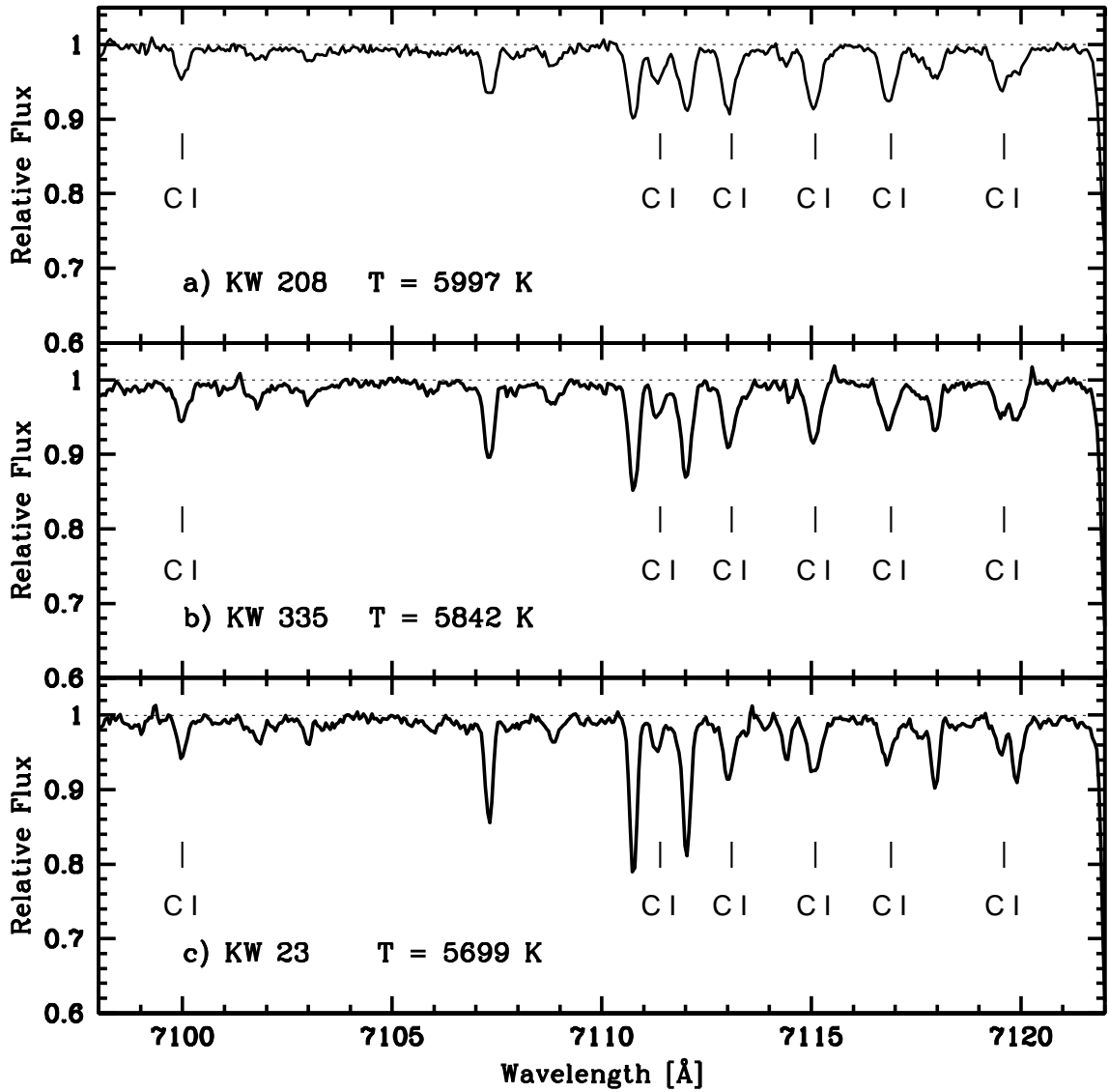}
\caption{Samples of our spectra in a 24 \AA\ region which contains several
high excitation C I lines.  Although the lines weaken with decreasing
temperature, they are not very sensitive to temperature.  The continuum is
shown as a light dotted line at 1.0.}
\end{figure}

\begin{figure}
\plotone{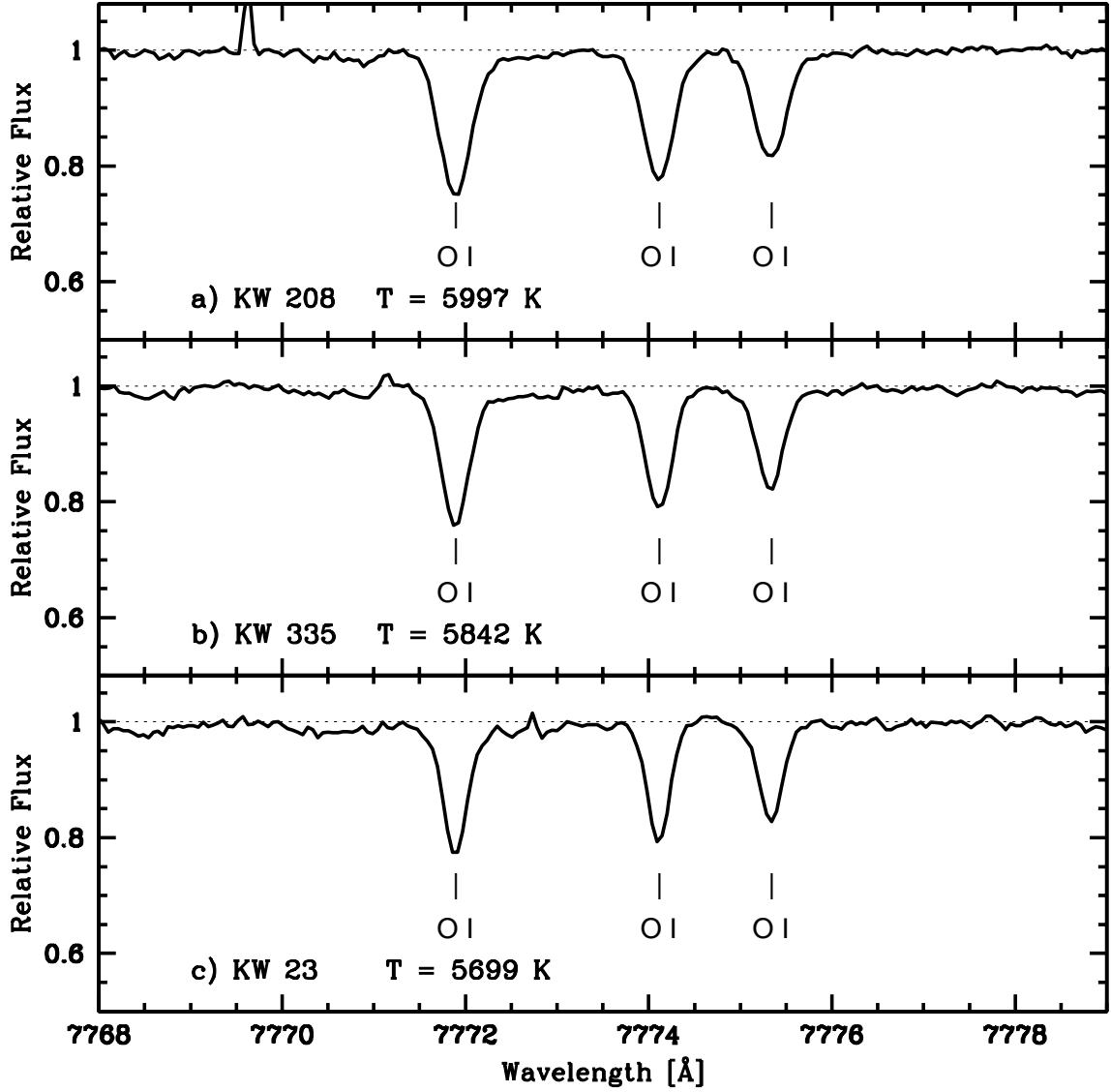}
\caption{Samples of our spectra in an 11 \AA\ region surrounding the O I
triplet.  The continuum is shown as a light dotted line at 1.0.}
\end{figure}

\begin{figure}
\plotone{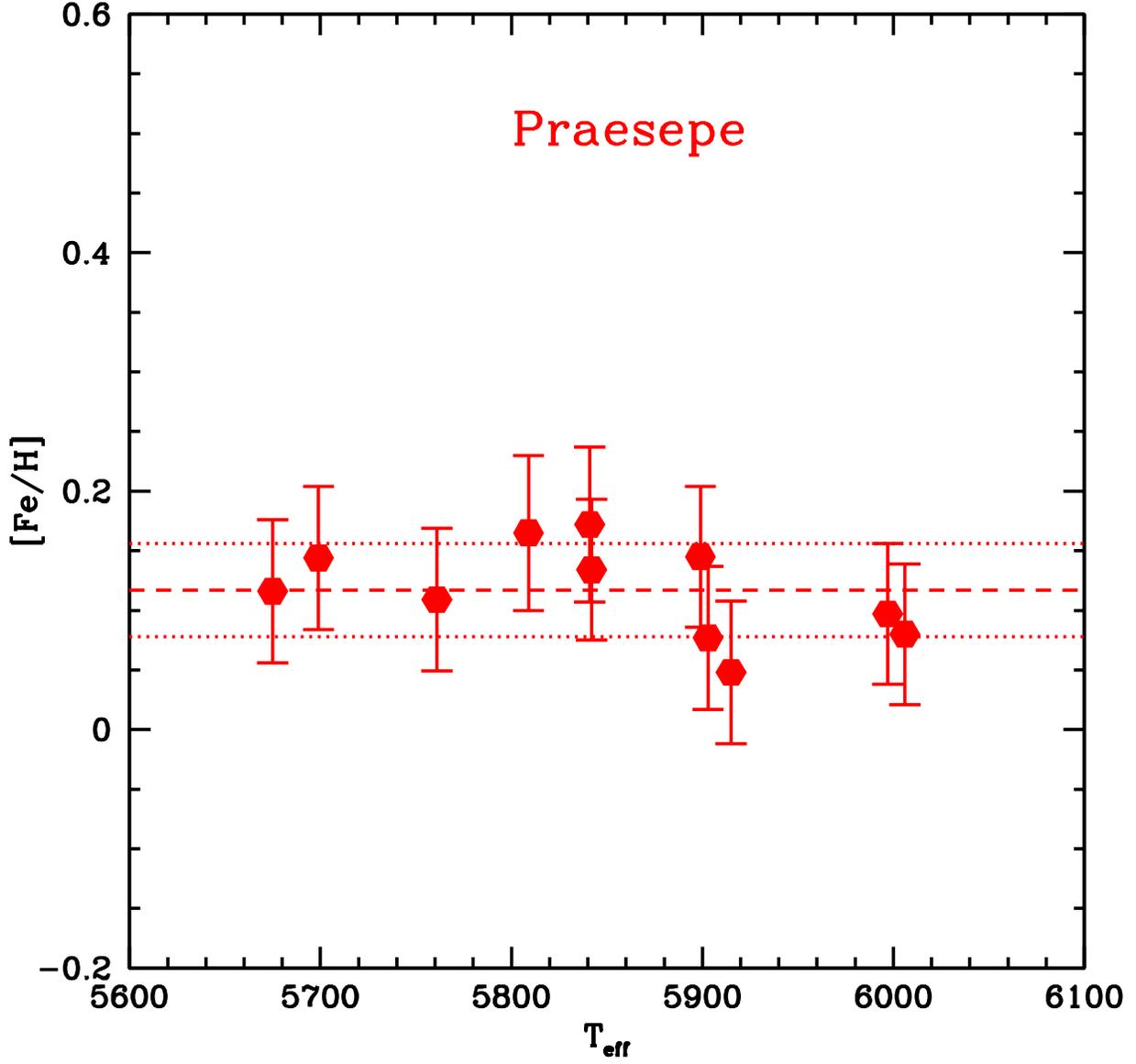}
\caption{The abundance of Fe in our Praesepe stars.  The error bars are those
due to the uncertainties in the stellar parameters.  The dotted line shows
the mean [Fe/H] = 0.117 $\pm$0.039.  The dashed lines are that 1$\sigma$
error.}
\end{figure}

\begin{figure}
\plotone{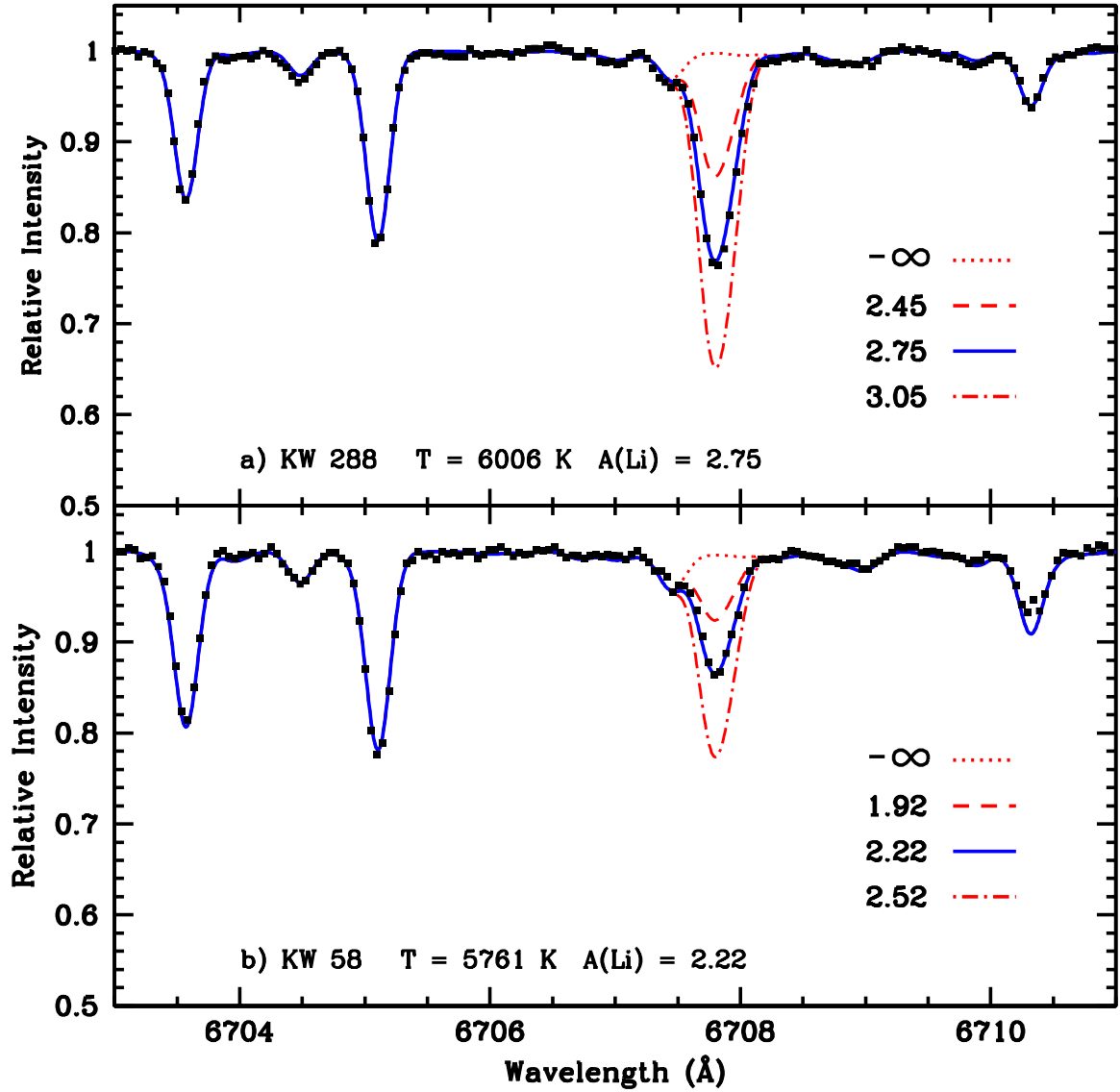}
\caption{Examples of the synthesized spectra in the Li regions for two
Praesepe stars.  The (black) squares are the observed spectra, the solid
(blue) line is the best synthetic fit, the dot-dash (red) line is a factor of
two more Li, the dashed (red) line is a factor of two less Li and the dotted
(red) shows the spectrum with no Li.  The other lines in this spectra region
are well matched by the synthesis.}
\end{figure}

\begin{figure}
\plotone{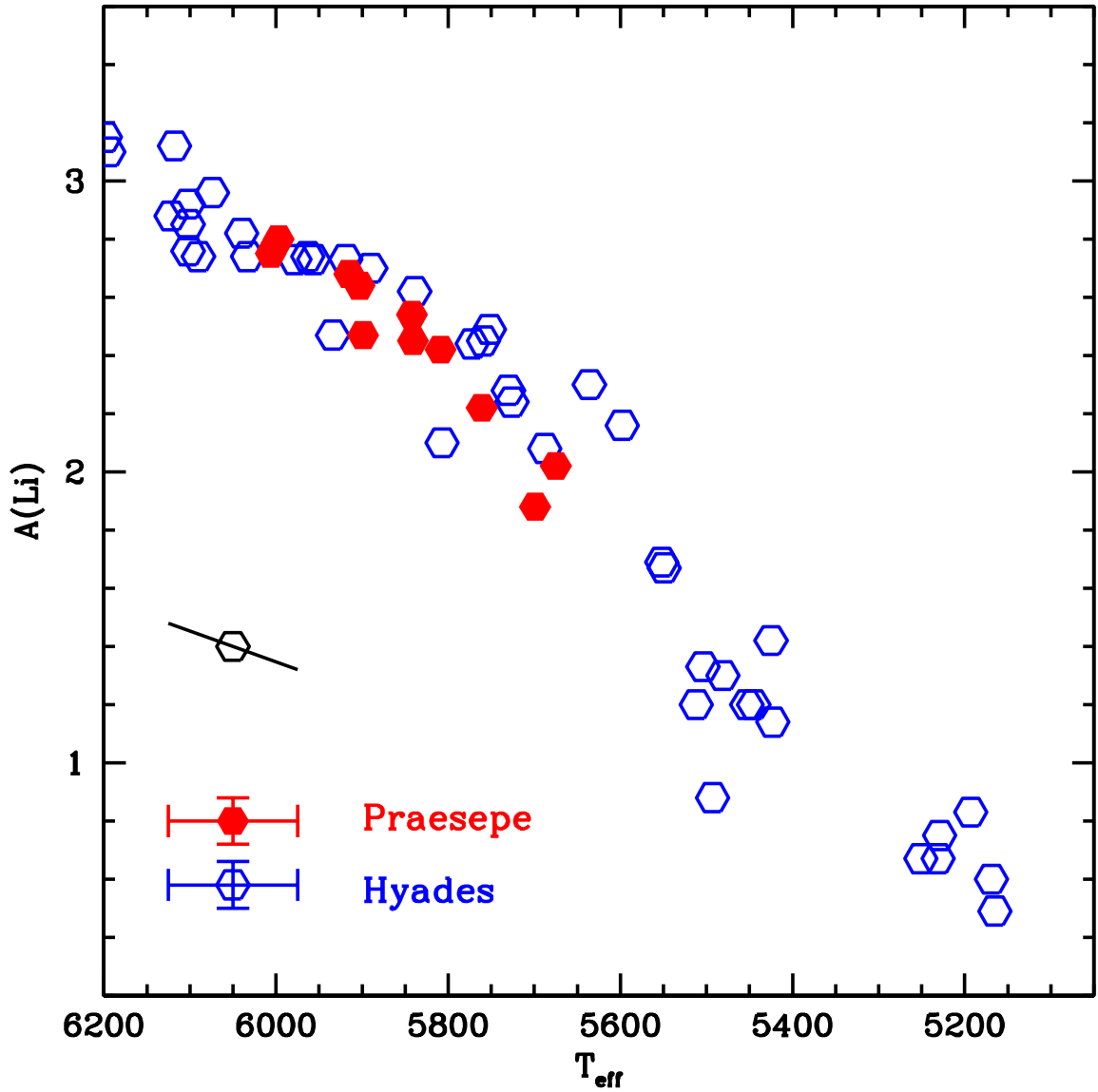}
\caption{Abundances of Li in the Hyades and Praesepe.  The Praesepe stars are
shown as filled hexagons while the Hyades stars are open hexagons.  A typical
uncertainty in both A(Li) and $T_{\rm eff}$ is shown in the lower left.  The
symbol above those error bars shows the direction and change in A(Li) due to
an uncertainty in T$_{\rm eff}$ of $\pm$75 K.}
\end{figure}

\begin{figure}
\plotone{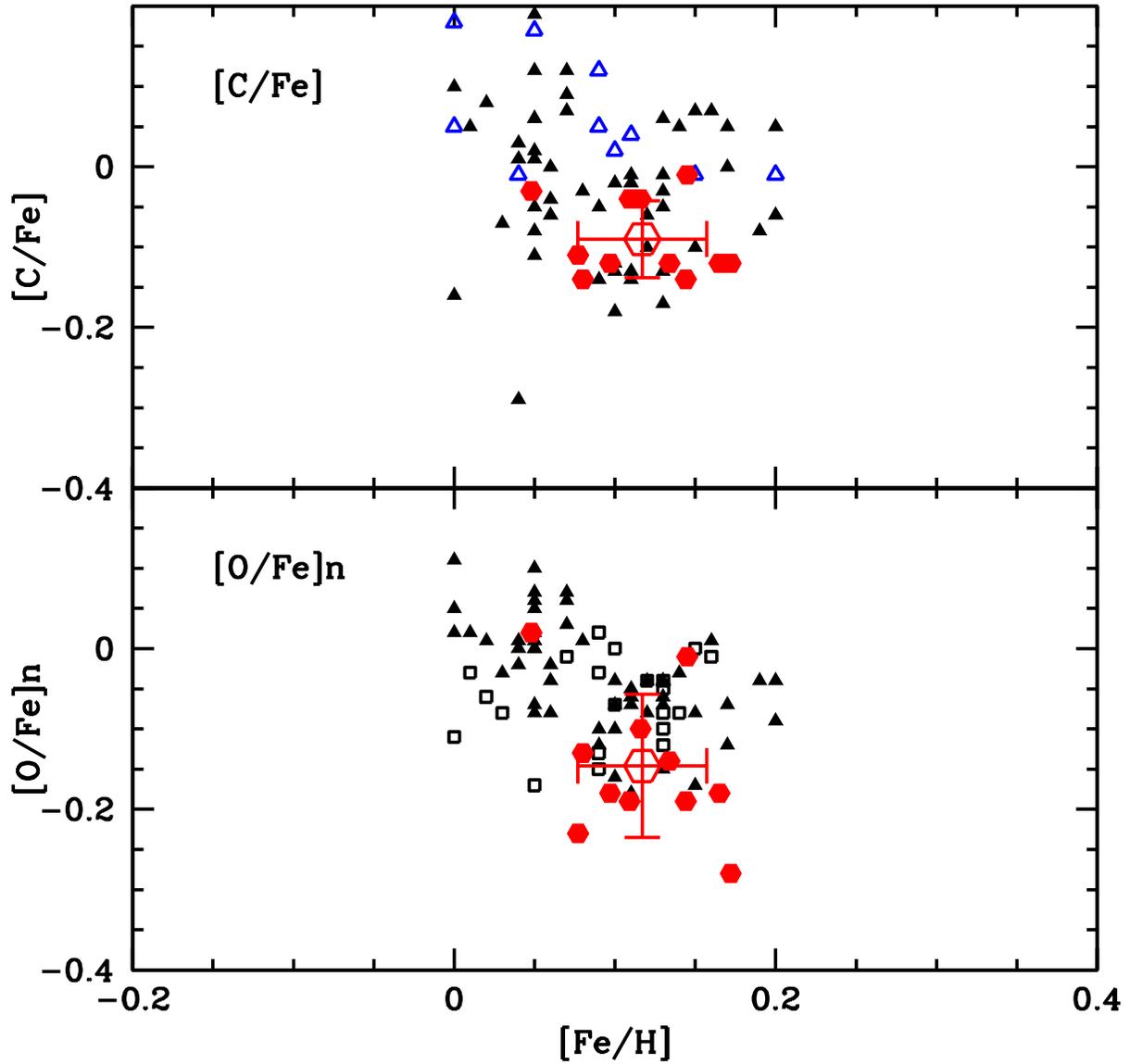}
\caption{Abundance ratios for C and O.  The Praesepe stars are the solid
hexagons (red) and the large open hexagon is the mean value for [X/Fe] with
the errorbars showing the standard deviation in the mean of the two
coordinates.  The Edvardsson et al.~(1993) abundances are the open squares,
the Reddy et al.~(2003, 2006) are the open triangles (blue), while the Takeda
\& Honda (2005) are filled triangles.}
\end{figure}

\begin{figure}
\plotone{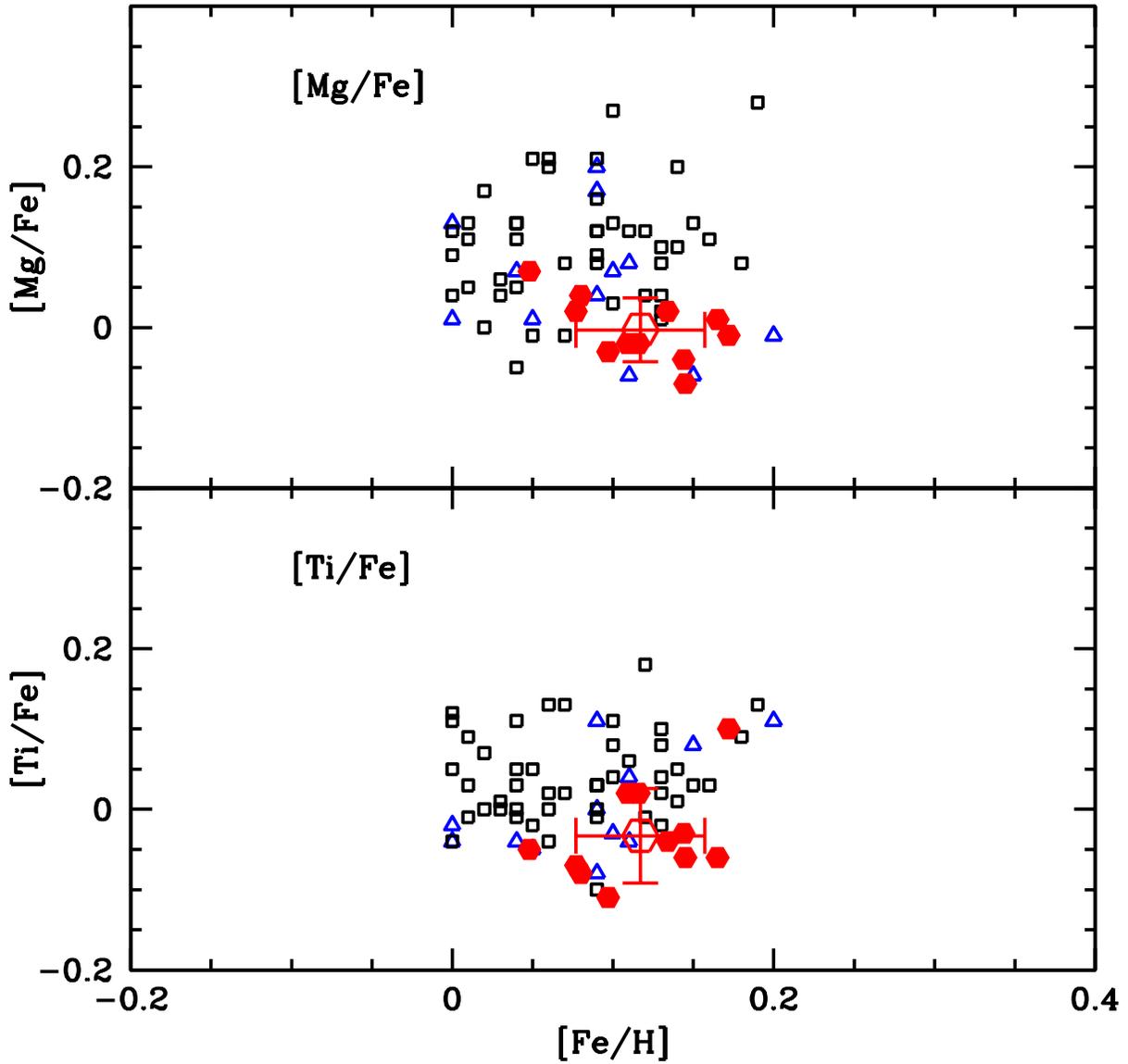}
\caption{Abundances of the alpha-elements, Mg and Ti, relative to Fe.  The
filled hexagons are the Praesepe stars and the large open hexagon is the mean
value for [X/Fe] with the errorbars showing the standard deviation in the mean
of the two coordinates.  The open squares are the selected sample of field
stars from Edvardsson et al.~(1993) and the open triangles (blue) are from the
two Reddy et al.~papers.}
\end{figure}

\begin{figure}
\plotone{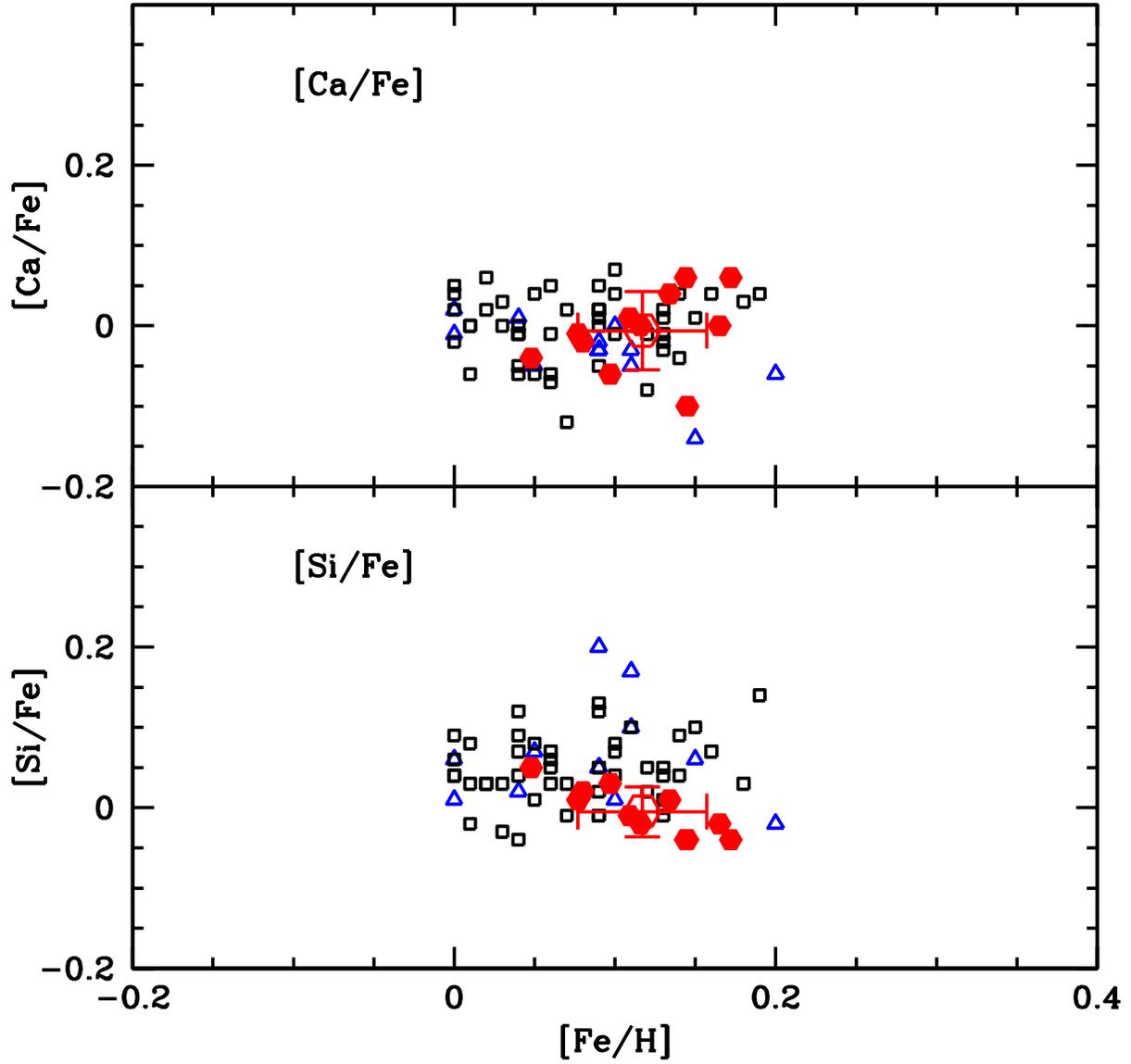}
\caption{Abundances of the alpha-elements, Ca and Si, relative to Fe.  The
Praesepe stars are the solid hexagons, the Edvardsson et al.~(1993) are the
open squares, the Reddy et al.~(2003, 2006) are the open triangles (blue).}
\end{figure}

\begin{figure}
\plotone{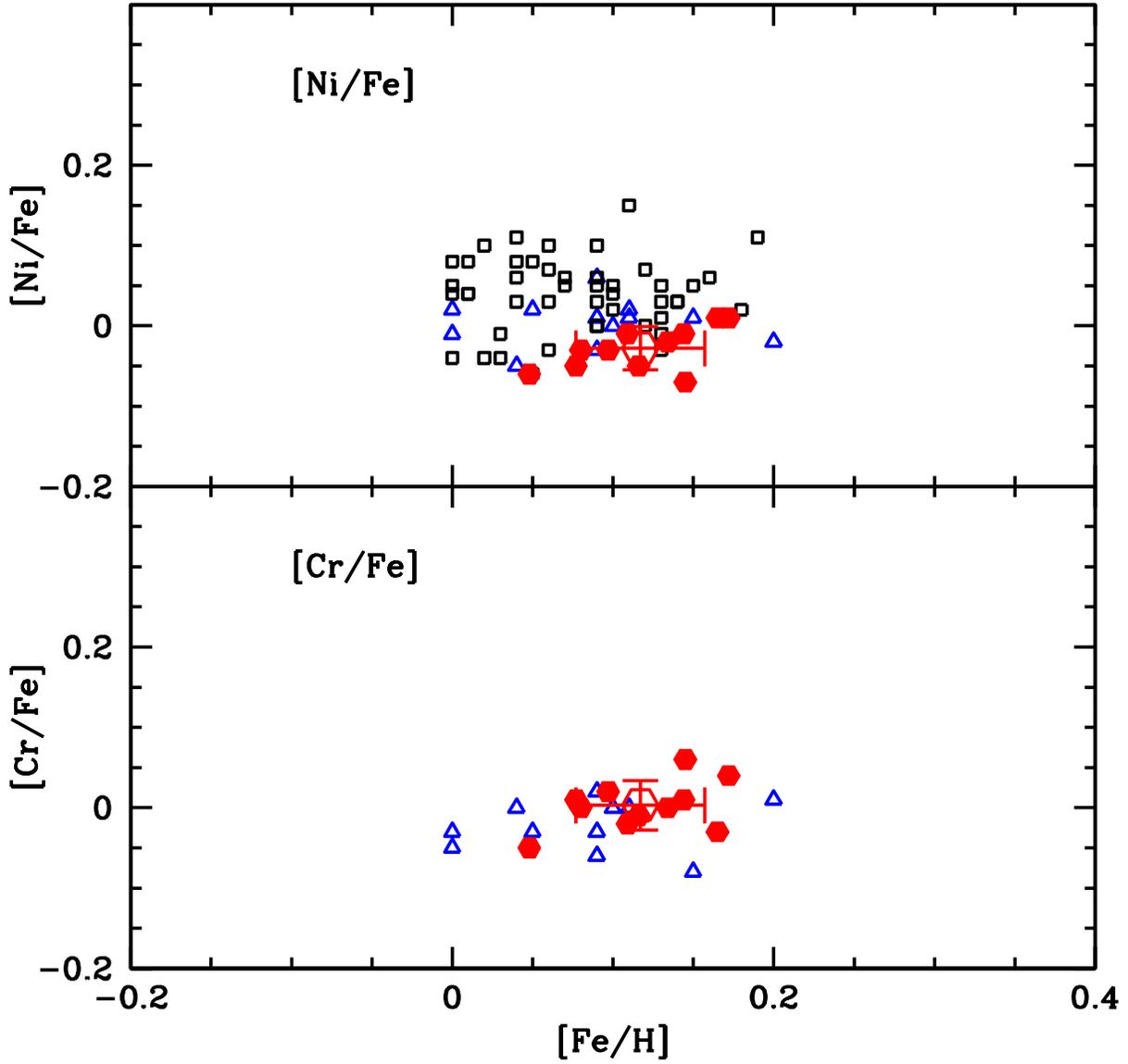}
\caption{Abundances of two Fe-peak elements, Cr and Ni relative to Fe.  The
Praesepe stars are the solid hexagons, the Edvardsson et al.~(1993) are the
open squares for Ni; they did not measure Cr.  The open triangles (blue) are
from the two Reddy et al.~papers.}
\end{figure}

\begin{figure}
\plotone{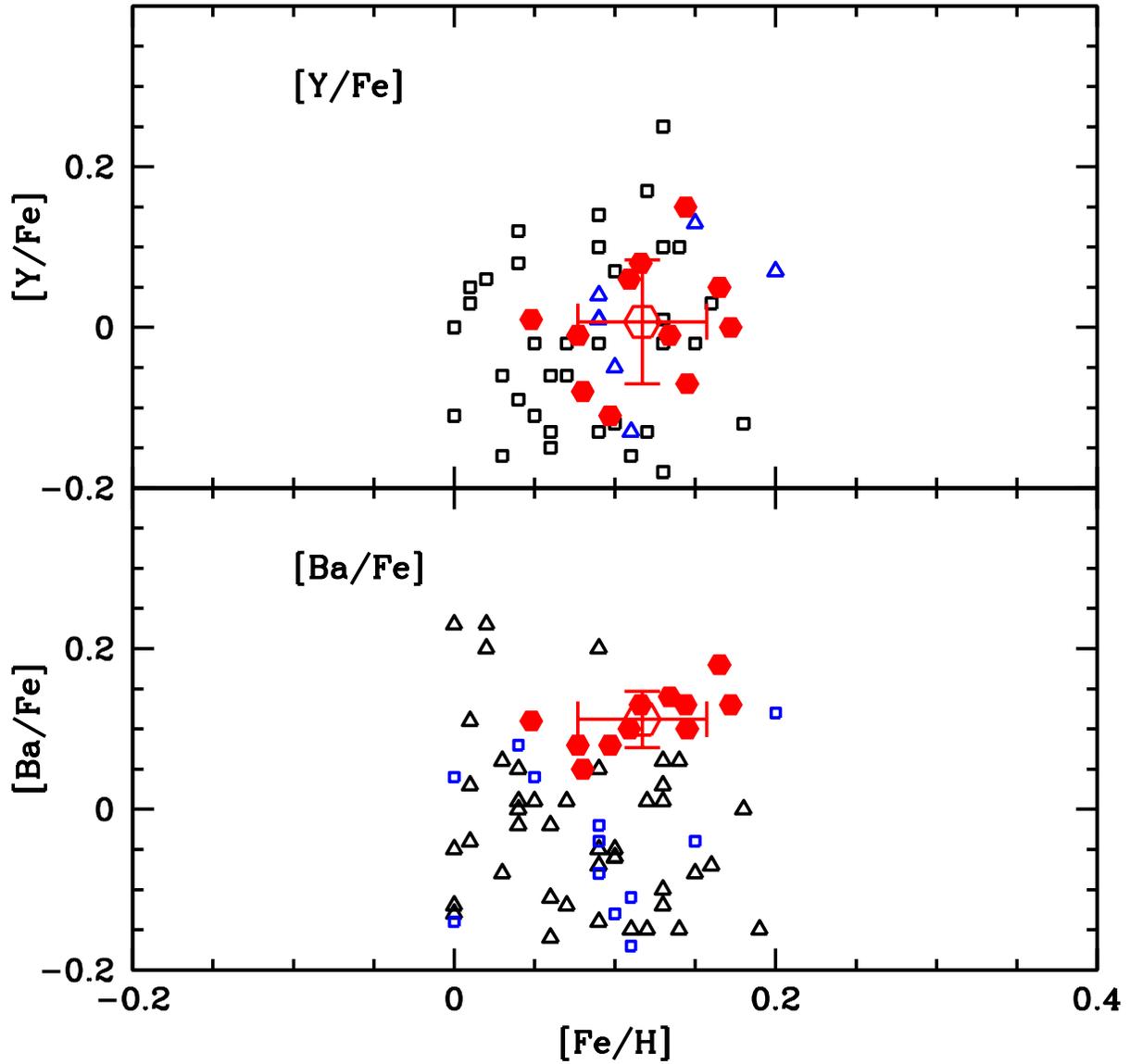}
\caption{Abundances of Y and Ba relative to Fe. Praesepe stars are the solid
hexagons; the Edvardsson et al.~(1993) are the open squares; the open
triangles (blue) are from the two Reddy et al.~papers.  Our Y results are in
accord with the field star results.  Our values for [Ba/Fe] at +0.11 $\pm$
0.04 are in the upper range of the field star values.}
\end{figure}

\begin{figure}
\plotone{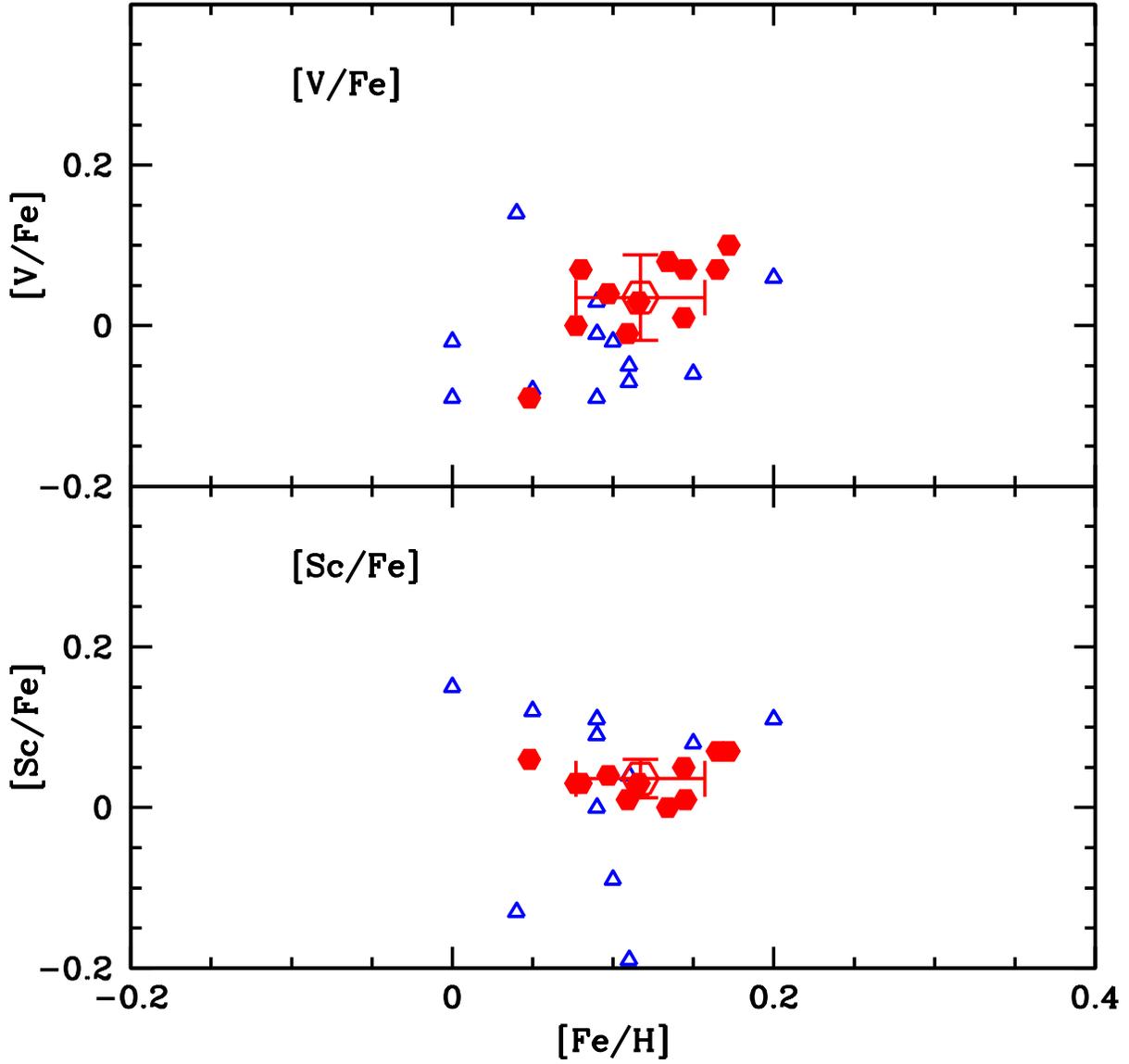}
\caption{Abundances of V and Sc relative to Fe.  The solid hexagons are the
Praesepe stars; open triangles are the field stars are from Reddy et
al.~(2003, 2006), Both samples are solar within the errors.}
\end{figure}

\begin{figure}
\plotone{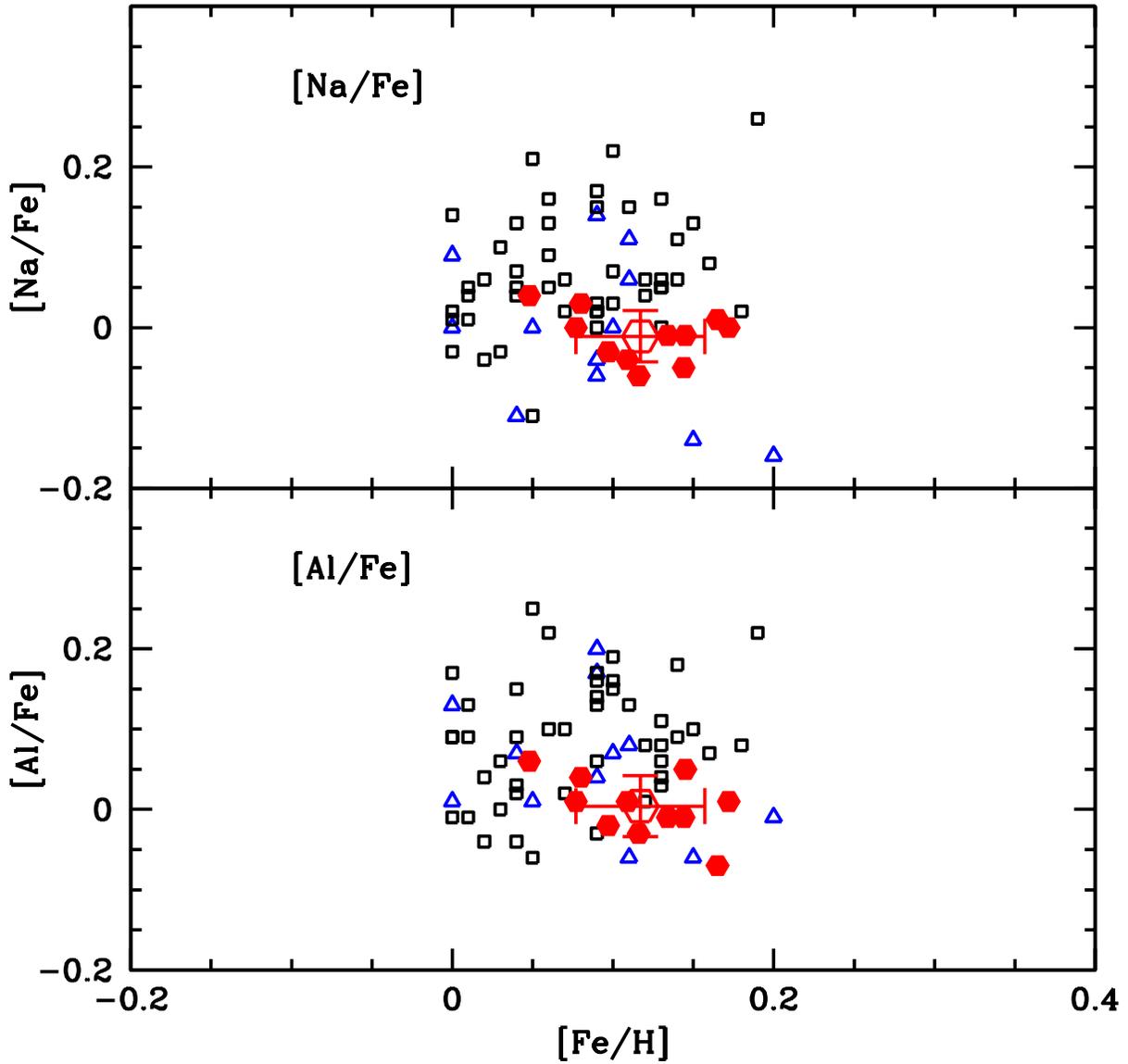}
\caption{Abundances of Al and Na relative to Fe.  The solid hexagons are the
Praesepe stars; the open triangles are the field stars are from Reddy et
al.~(2003, 2006), open squares are from Edvardsson et al.~(1993).  Both Al and
Na are solar which is generally lower than the filed stars.}
\end{figure}

\begin{figure}
\plotone{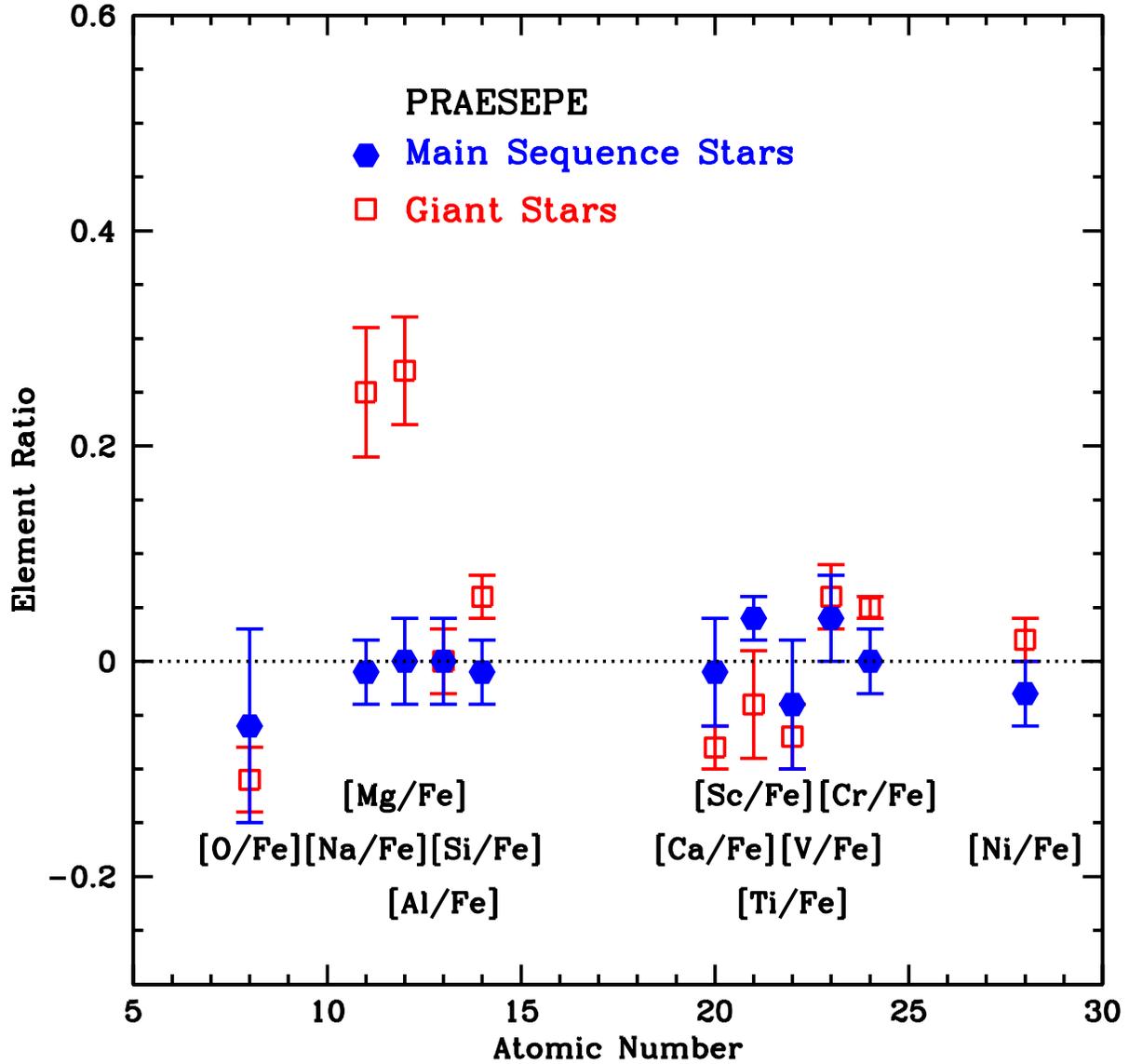}
\caption{Abundance averages in our 11 main sequence stars compared with the
abundance average in the three red giants of Carrera \& Pancino (2011).  Our
dwarfs with 1 sigma error bars are the (blue) hexagons while the results for
the giants are shown by the (red) open squares.  The horizontal line
represents the solar values.  With the exception of Na and Mg the agreement is
very good.  See text for discussion.}
\end{figure}

\clearpage
\input{line.eqw.tex}


\end{document}

%% file: tab1.tex

\singlespace
\begin{center}
\begin{deluxetable}{lrrcrcc} 
\tablewidth{0pc}
\tablenum{1}
\tablecolumns{7} 
\tablecaption{Keck/HIRES Observations of Praesepe} 
\tablehead{ 
\colhead{Star}  &  \colhead{V}  & \colhead{B-V}  
& \colhead{ref.\tablenotemark{1}} &
\colhead{Night} & \colhead{Exp. Time} & \colhead{Total}  \\
\colhead{} & \colhead{} & \colhead{} & \colhead{} & \colhead{UT} 
& \colhead{(min)} 
& \colhead{S/N}
} 
\startdata 
KW 23  & 11.29 & 0.710 & 1 & 11 Jan 2003  & \phn 25 & 149 \\
KW 30  & 11.40 & 0.730 & 1 & 11 Jan 2003  & \phn 25 & 141 \\
KW 58  & 11.26 & 0.671 & 2 & 11 Feb 2003  & \phn 25 & 159 \\
KW 181 & 10.47 & 0.588 & 2 & 11 Jan 2003  & \phn 10 & \phn93  \\
KW 208 & 10.66 & 0.583 & 2 & 11 Jan 2003  & \phn 14 & 153 \\
KW 288 & 10.69 & 0.593 & 2 & 11 Feb 2003  & \phn 15 & 150 \\
KW 301 & 11.17 & 0.655 & 2 & 11 Feb 2003  & \phn 25 & 155 \\
KW 335 & 11.03 & 0.651 & 2 & 11 Jan 2003  & \phn 20 & 156 \\
KW 399 & 11.93 & 0.624 & 2 & 11 Jan 2003  & \phn 20 & 166 \\
KW 432 & 11.05 & 0.646 & 2 & 11 Feb 2003  & \phn 20 & 143 \\
KW 508 & 10.77 & 0.594 & 2 & 11 Jan 2003  & \phn 20 & 159 \\
\enddata 

\tablenotetext{1}{
1 = Mendoza (1967),
2 = Johnson (1952)
}

\end{deluxetable} 
\end{center}


%% file: tab2.tex

\singlespace
\begin{center}
\begin{deluxetable}{lrrrrrrrrr}
\tablewidth{0pc}
\tablecolumns{10} 
\tablenum{2}
\tablecaption{Colors}
\tablehead{
\colhead{Star}  & \colhead{B} &  \colhead{V} &  \colhead{J} &  \colhead{H} &
 \colhead{K} &  \colhead{V$_c$} &  \colhead{V-R$_c$} 
&  \colhead{R$_c$$-$I$_c$} & \colhead{V$-$I$_c$} 
}

\startdata 
KW 23  & 12.00 & 11.29 & 10.075 & 9.780 & 9.686 & \nodata & \nodata & \nodata & \nodata \\
KW 30  & 12.13 & 11.40 & 10.187 & 9.898 & 9.803 & \nodata & \nodata & \nodata & \nodata \\
KW 58  & 11.93 & 11.26 & 10.079 & 9.791 & 9.689 & \nodata & \nodata & \nodata & \nodata \\
KW 181 & 11.06 & 10.47 &  9.357 & 9.088 & 8.997 & 10.488  & 0.339   & 0.325
& 0.664 \\
KW 208 & 11.24 & 10.66 &  9.565 & 9.357 & 9.259 & \nodata & \nodata & \nodata & \nodata \\
KW 288 & 11.28 & 10.69 &  9.640 & 9.401 & 9.336 & 10.698  & 0.333   & 0.308
& 0.641 \\
KW 301 & 11.83 & 11.17 & 10.012 & 9.698 & 9.655 & \nodata & \nodata & \nodata & \nodata \\
KW 335 & 11.68 & 11.03 &  9.864 & 9.588 & 9.507 & \nodata & \nodata & \nodata & \nodata \\
KW 399 & 11.55 & 10.93 &  9.812 & 9.549 & 9.469 & \nodata & \nodata & \nodata & \nodata \\
KW 432 & 11.70 & 11.05 &  9.869 & 9.627 & 9.544 & \nodata & \nodata & \nodata & \nodata \\
KW 508 & 11.36 & 10.77 &  9.659 & 9.416 & 9.359 & 10.761  & 0.334   & 0.326
& 0.660 \\
\enddata 

\end{deluxetable} 
\end{center}


%% file: tab3.tex

\singlespace
\begin{center}
\begin{deluxetable}{lrrrrrrcr}
\tablewidth{0pc}
\tablecolumns{9} 
\tablenum{3}
\tablecaption{Temperatures}
\tablehead{
\colhead{Star}  & \colhead{T$_{B-V}$} &  \colhead{T$_{V-J}$} &  
\colhead{T$_{V-H}$} &  \colhead{T$_{V-Ic}$} &
 \colhead{T$_{V-Rc}$} & \colhead{T$_{Rc-Ic}$}
& \colhead{mean T$_{\rm eff}$} & \colhead{$\sigma$}
}
\startdata 
KW 23  & 5615 &	5703 & 5772 & \nodata & \nodata & \nodata & 5697 & $\pm$79 \\
KW 30  & 5554 &	5708 & 5770 & \nodata & \nodata & \nodata & 5677 & $\pm$111 \\
KW 58  & 5697 &	5771 & 5813 & \nodata & \nodata & \nodata & 5761 & $\pm$59 \\
KW 181 & 5922 &	5910 & 5954 & 5868    & 5918    & 5824    & 5899 & $\pm$46 \\
KW 208 & 5978 &	5950 & 6065 & \nodata & \nodata & \nodata & 5997 & $\pm$60 \\
KW 288 & 6006 &	6050 & 6078 & 5958    & 5954    & 5988    & 6006 & $\pm$50 \\
KW 301 & 5776 &	5815 & 5836 & \nodata & \nodata & \nodata & 5809 & $\pm$30 \\
KW 335 & 5861 &	5800 & 5867 & \nodata & \nodata & \nodata & 5842 & $\pm$37 \\
KW 399 & 5874 &	5901 & 5932 & \nodata & \nodata & \nodata & 5903 & $\pm$29 \\
KW 432 & 5841 &	5768 & 5914 & \nodata & \nodata & \nodata & 5841 & $\pm$73 \\
KW 508 & 5958 &	5919 & 5951 & 5887    & 5943    & 5831    & 5915 & $\pm$48 \\
\enddata 

\end{deluxetable} 
\end{center}


%% file: tab4.tex

\singlespace
\begin{center}
\begin{deluxetable}{lccccccc}
\tablenum{4}
\tablewidth{0pc}
\tablecolumns{8} 
\tablecaption{[Fe/H] Abundances from Fe I and Fe II.}
\tablehead{
\colhead{Star}  & \colhead{[FeI/H]} &  \colhead{$\sigma$} & \colhead{num.} & 
\colhead{[FeII/H]} &  \colhead{$\sigma$} & \colhead{num.} & 
\colhead{mean [Fe/H]} 
}
\startdata 
KW 23  & 0.15  & $\pm$0.08  & 54 & 0.09 & $\pm$0.11 & 6 & 0.144 \\
KW 30  & 0.12  & $\pm$0.08  & 55 & 0.08 & $\pm$0.09 & 6 & 0.116 \\
KW 58  & 0.11  & $\pm$0.09  & 55 & 0.10 & $\pm$0.06 & 6 & 0.109 \\
KW 181 & 0.13  & $\pm$0.10  & 50 & 0.29 & $\pm$0.09 & 5 & 0.145 \\
KW 208 & 0.10  & $\pm$0.10  & 55 & 0.07 & $\pm$0.09 & 6 & 0.097 \\
KW 288 & 0.08  & $\pm$0.09  & 54 & 0.08 & $\pm$0.07 & 6 & 0.080 \\
KW 301 & 0.17  & $\pm$0.09  & 55 & 0.12 & $\pm$0.09 & 6 & 0.165 \\
KW 335 & 0.14  & $\pm$0.09  & 55 & 0.08 & $\pm$0.09 & 6 & 0.134 \\
KW 399 & 0.08  & $\pm$0.09  & 55 & 0.05 & $\pm$0.09 & 6 & 0.077 \\
KW 432 & 0.18  & $\pm$0.09  & 55 & 0.10 & $\pm$0.09 & 6 & 0.172 \\
KW 508 & 0.04  & $\pm$0.10  & 55 & 0.13 & $\pm$0.06 & 6 & 0.048 \\
\enddata 

\end{deluxetable} 
\end{center}


%% file: tab5.tex

\singlespace
\begin{center}
\begin{deluxetable}{lcccc}
\tablewidth{0pc}
\tablecolumns{5} 
\tablenum{5}
\tablecaption{ Parameters for the Stellar Models}
\tablehead{
\colhead{Star}  & \colhead{$T_{\rm eff}$} &  \colhead{log g} &
\colhead{[Fe/H]} & \colhead{$\xi$} 
}
\startdata 
KW 23  & 5699 & 4.43 & 0.12 & 1.10 \\
KW 30  & 5675 & 4.44 & 0.12 & 1.07 \\
KW 58  & 5761 & 4.42 & 0.12 & 1.16 \\
KW 181 & 5899 & 4.39 & 0.12 & 1.31 \\
KW 208 & 5997 & 4.38 & 0.12 & 1.40 \\
KW 288 & 6006 & 4.38 & 0.12 & 1.41 \\
KW 301 & 5809 & 4.41 & 0.12 & 1.21 \\
KW 335 & 5842 & 4.40 & 0.12 & 1.26 \\
KW 399 & 5903 & 4.40 & 0.12 & 1.31 \\
KW 432 & 5841 & 4.40 & 0.12 & 1.25 \\
KW 508 & 5915 & 4.39 & 0.12 & 1.33 \\
\enddata 

\end{deluxetable} 
\end{center}


%% file: tab6.tex
\tightenlines
\singlespace
\begin{center}
\begin{deluxetable}{lrrrrrrrrrrrrr}
\tabletypesize{\small}
\rotate
\thispagestyle{empty}
\tablewidth{0pc}
\tablecolumns{14} 
\tablenum{6}
\tablecaption{Chemical Abundances for Praesepe Stars}
\tablehead{
\colhead{} & \colhead{KW 23} & \colhead{KW 30} & \colhead{KW 58} 
& \colhead{KW 181} & \colhead{KW 208} 
& \colhead{KW 288} & \colhead{KW 301} & \colhead{KW 335} & \colhead{KW 399} &
\colhead{KW 432} & \colhead{KW 508} & \colhead{mean} &\colhead{$\sigma$} 
}
\startdata 
$[$Fe/H$]$  & 0.144 & 0.116 & 0.109  & 0.145  & 0.097  & 0.080  & 0.163  & 0.134 &0.077  & 0.172 & 0.048 & 0.117 & 0.039 \\
A(Li) & 1.88 & 2.02 & 2.22 & 2.47 & 2.80 & 2.75 & 2.42 & 2.54 & 2.64 & 2.45 &
2.68 & \nodata & \nodata \\
$[$C/Fe$]$  & $-$0.14 & $-$0.04 & $-$0.04 & $-$0.01 & $-$0.12 & $-$0.14 & $-$0.12 & $-$0.12 & $-$0.11 & $-$0.12 & $-$0.03 & $-$0.090 & 0.048 \\
$[$O/Fe$]$n & $-$0.19 & $-$0.10 & $-$0.19 &$-$0.01 & $-$0.18 & $-$0.13 & $-$0.18 &
$-$0.14 & $-$0.23 & $-$0.28 & 0.02 & $-$0.146 & 0.089 \\
$[$Na/Fe$]$ & $-$0.05 & $-$0.06 & $-$0.04 & $-$0.01 & $-$0.03 & 0.03 & 0.01 &
$-$0.01 & 0.00 & 0.00 & 0.04 & $-$0.011 & 0.032 \\
$[$Mg/Fe$]$ & $-$0.04 & $-$0.02 & $-$0.02 & $-$0.07 & $-$0.03 & 0.04 & 0.01 &
0.02 & 0.02 & $-$0.01 & 0.07  & $-$0.003 & 0.040 \\
$[$Al/Fe$]$ & $-$0.01 & $-$0.03 & 0.01 & 0.05 & $-$0.02 & 0.04 & $-$0.07 & $-$0.01 & 0.01 & 0.01 & 0.06 & 0.004 & 0.038 \\
$[$Si/Fe$]$ & $-$0.04 & $-$0.02 & $-$0.01 & $-$0.04 & 0.03 & 0.02 & $-$0.02 &
0.01 & 0.01 & $-$0.04 & 0.05 & $-$0.005 & 0.031 \\
$[$Ca/Fe$]$ & 0.06 & 0.00 & 0.01 & $-$0.01 & $-$0.06 & $-$0.02 & 0.00 & 0.04 &
$-$0.01 & 0.06 & $-$0.04 & $-$0.006 & 0.049 \\
$[$Sc/Fe$]$ & 0.05 & 0.03 & 0.01 & 0.01 & 0.04 & 0.03 & 0.07 & 0.00 & 0.03 &
0.07 & 0.06 & 0.036 & 0.024 \\
$[$Ti/Fe$]$ & $-$0.03 & 0.02 & 0.02 & $-$0.06 & $-$0.11 & $-$0.08 & $-$0.06 &
$-$0.04 & $-$0.07 & 0.10 & $-$0.05 & $-$0.033 & 0.059 \\
$[$V/Fe$]$  & $-$0.01 & 0.03 & 0.02 & 0.04 & 0.06 & 0.11 & 0.02 & 0.07 & 0.04 & 0.05 & $-$0.02 & 0.037 & 0.036 \\
$[$Cr/Fe$]$ & 0.01 & $-$0.01 & $-$0.02 & 0.06 & 0.02 & 0.00 & $-$0.03 & 0.00 & 0.01 &
0.04 & $-$0.05 & 0.003 & 0.031 \\
$[$Ni/Fe$]$ & $-$0.01 & $-$0.05 & $-$0.01 & $-$0.07 & $-$0.03 & $-$0.03 & 0.01
& $-$0.02 & $-$0.05 & 0.01 & $-$0.06 & $-$0.028 & 0.027 \\
$[$Y/Fe$]$  & 0.15 & 0.08 & 0.06 & $-$0.07 & $-$0.11 & $-$0.08 & 0.05  &
$-$0.01 & $-$0.01 & 0.00 & 0.01 & 0.007 & 0.077 \\
$[$Ba/Fe$]$ & 0.13 & 0.13 & 0.10 & 0.10 & 0.08 & 0.05 & 0.18 & 0.14 & 0.08
&0.13 & 0.11 & 0.112 & 0.035 \\
\enddata 

\end{deluxetable} 
\end{center}



%% file: tab7.tex
\tightenlines

\singlespace
\begin{center}
\begin{deluxetable}{lccrcc}
\tablenum{7}
\tablewidth{0pc}
\tablecolumns{6} 
\tablecaption{Errors on Fe I Determination due to Parameter Uncertainties}
\tablehead{
\multicolumn{1}{c}{Star}  & 
\multicolumn{1}{c}{$\Delta$T$_{\rm eff}$} &
\multicolumn{1}{c}{$\Delta$log g} &
\multicolumn{1}{c}{$\Delta$[Fe/H]} & 
\multicolumn{1}{c}{$\Delta\xi$} & 
\multicolumn{1}{c}{Total} \\
\multicolumn{1}{c}{} & 
\multicolumn{1}{c}{$\pm$75 K} & 
\multicolumn{1}{c}{$\pm$0.20} & 
\multicolumn{1}{c}{$\pm$0.10} & 
\multicolumn{1}{c}{$\pm$0.20} & 
\multicolumn{1}{c}{}
}
\startdata
KW 23  & $\pm$0.05 & $\mp$0.01  & $\pm$0.01 & $\pm$0.03 & 0.060 \\
KW 30  & $\pm$0.04 & $\mp$0.02  & 0.00	    & $\pm$0.04 & 0.060 \\
KW 58  & $\pm$0.05 & $\mp$0.01  & $\pm$0.01 & $\pm$0.04 & 0.060 \\
KW 181 & $\pm$0.05 & $\mp$0.01  & 0.00	    & $\pm$0.01 & 0.059 \\
KW 208 & $\pm$0.05 & $\mp$0.01  & 0.00	    & $\pm$0.03 & 0.059 \\
KW 288 & $\pm$0.04 & $\mp$0.01  & 0.00	    & $\pm$0.05 & 0.059 \\
KW 301 & $\pm$0.05 & $\mp$0.01  & 0.00	    & $\pm$0.04 & 0.065 \\
KW 335 & $\pm$0.05 & $\mp$0.01  & 0.00	    & $\pm$0.03 & 0.059 \\
KW 399 & $\pm$0.05 & $\mp$0.01  & $\pm$0.01 & $\pm$0.03 & 0.060 \\
KW 432 & $\pm$0.05 & $\mp$0.01  & 0.00	    & $\pm$0.04 & 0.065 \\
KW 508 & $\pm$0.05 & $\mp$0.01  & 0.00      & $\pm$0.01 & 0.060 \\

\enddata 

\end{deluxetable} 
\end{center}


%% file: tab8.tex
\tightenlines

\singlespace
\begin{center}
\begin{deluxetable}{llcrrrc}
\tablenum{8}
\tablewidth{0pc}
\tablecolumns{8} 
\tablecaption{Errors due to Uncertainties in the Stellar Parameters}
\tablehead{
\multicolumn{1}{c}{Star}  & 
\multicolumn{1}{c}{Element}  & 
\multicolumn{1}{c}{$\Delta$T$_{\rm eff}$} &
\multicolumn{1}{c}{$\Delta$log g} &
\multicolumn{1}{c}{$\Delta$[Fe/H]} & 
\multicolumn{1}{c}{$\Delta\xi$} & 
\multicolumn{1}{c}{Total} \\
\multicolumn{1}{c}{} & 
\multicolumn{1}{c}{} & 
\multicolumn{1}{c}{$\pm$75 K} & 
\multicolumn{1}{c}{$\pm$0.20} & 
\multicolumn{1}{c}{$\pm$0.10} & 
\multicolumn{1}{c}{$\pm$0.20} & 
\multicolumn{1}{c}{}
}
\startdata
KW 23 & C I  & $\mp$0.06 & $\pm$0.07 & $\mp$0.01 & 0.00       & 0.093  \\
&	O I  & $\mp$0.08 & $\pm$0.05 & 0.00     & $\mp$0.01  & 0.095 \\
&	Na I & $\pm$0.07 & 0.00      & $\pm$0.03 & $\pm$0.02 & 0.079 \\
&	Mg I & $\pm$0.03 & $\mp$0.02 & 0.00 & $\pm$0.01	 & 0.037 \\
&	Al I & $\pm$0.04 & $\mp$0.01 & 0.00     &	$\mp$0.01 & 0.042  \\
&	Si I & $\pm$0.01 & 0.00	   & $\pm$0.02 & $\mp$0.01 & 0.024 \\
&	Ca I & $\pm$0.06 & $\mp$0.04 & $\pm$0.01 & $\mp$0.04 & 0.083 \\
&	Sc II& 0.00      & $\pm$0.09 & $\pm$0.03 & $\mp$0.02 & 0.097 \\
&	Ti I & $\pm$0.07 & $\mp$0.01 & 0.00      & $\mp$0.01 & 0.071 \\
&	V I  & $\pm$0.09 & 0.00      & 0.00      & $\mp$0.01 & 0.091 \\
&	Cr I & $\pm$0.06 & $\mp$0.01 & $\pm$0.01 & $\mp$0.02 & 0.065 \\
&	Ni I & $\pm$0.04 & $\mp$0.01 & $\pm$0.01 & $\mp$0.05 & 0.066 \\
&	Y I  & $\pm$0.11 & $\mp$0.01 & 0.00      & 0.00      & 0.110 \\
&       Y II & $\mp$0.01 & $\pm$0.09 & $\mp$0.02 & $\mp$0.02 & 0.095 \\
&	Ba II& $\pm$0.03 & $\pm$0.05 & $\mp$0.11  & $\mp$0.11 & 0.166  \\
\hline
KW 208 & C I & $\mp$0.04  & $\pm$0.05 & $\mp$0.01  & $\mp$0.01  & 0.066 \\
& 	 O I & $\mp$0.05  & $\pm$0.03 & 0.00	& $\mp$0.02 & 0.062 \\
&	Na I & $\pm$0.06  & 0.00      & $\pm$0.02 & $\pm$0.01 & 0.064 \\
& 	Mg I & $\pm$0.03  & $\mp$0.01 & $\pm$0.01  & 0.00   & 0.037  \\
& 	Al I & $\pm$0.04  & 0.00      & 0.00   & 0.00 & 0.040 \\
& 	Si I & $\pm$0.02  & $\mp$0.01  & $\pm$0.01  & $\mp$0.01  & 0.026 \\
& 	Ca I & $\pm$0.06  & $\mp$0.04  & 0.00  & $\mp$0.04 & 0.082 \\
& 	Sc II& $\pm$0.01  & $\pm$0.07  & $\pm$0.03  & $\mp$0.02  & 0.079 \\
& 	Ti I & $\pm$0.07  & 0.00  & 0.00 & $\mp$0.01 & 0.071 \\
& 	V I  & $\pm$0.08  & 0.00  & 0.00  & $\mp$0.01  & 0.081 \\
& 	Cr I & $\pm$0.04  & $\mp$0.02 & $\mp$0.01  & $\mp$0.02 & 0.050 \\
& 	Ni I & $\pm$0.05  & $\mp$0.01  & 0.00  & $\mp$0.04  & 0.065  \\
& 	Y I  & $\pm$0.10  & 0.00       & 0.00   & 0.00  & 0.100 \\
&       Y II & 0.00       & $\pm$0.08  & $\pm$0.03 & $\mp$0.01  & 0.086 \\
& 	Ba II& $\pm$0.03  & $\pm$0.06  & $\pm$0.04  & $\mp$0.09  & 0.119   \\
KW 335 & C I & $\mp$0.06  & $\pm$0.06  & $\mp$0.01  & $\mp$0.01 & 0.086  \\
&	O I  & $\mp$0.06  & $\pm$0.04  & $\pm$0.01  & $\mp$0.01 & 0.074  \\
&	Na I & $\pm$0.07  & $\pm$0.01  & $\pm$0.02 & $\pm$0.03 & 0.079 \\
&	Mg I & $\pm$0.03  & $\mp$0.02  & 0.00  & 0.00 & 0.036 \\
&	Al I & $\pm$0.03  & $\mp$0.01  & 0.00  & $\mp$0.01 & 0.033 \\
&	Si I & $\pm$0.02  & $\mp$0.01  & $\pm$0.02  & $\mp$0.01 & 0.032 \\
&	Ca I & $\pm$0.05  & $\mp$0.04  & 0.00  & $\mp$0.04 & 0.076 \\ 
&	Sc II& 0.00       & $\pm$0.09  & $\pm$0.04  & $\mp$0.02 & 0.100 \\    
&	Ti I & $\pm$0.07  & 0.00       & 0.00  & $\mp$0.01 & 0.071 \\ 
&	V I  & $\pm$0.07  & $\mp$0.01  & 0.00  & $\mp$0.02 & 0.074 \\ 
&	Cr I & $\pm$0.06  & $\mp$0.01  & 0.00  & $\mp$0.02 & 0.064 \\ 
&	Ni I & $\pm$0.05  & $\mp$0.01  & $\pm$0.01  & $\mp$0.04 & 0.066 \\ 
&	Y I  & $\pm$0.10  & 0.00       & 0.00  & 0.00 & 0.100 \\ 
&	Y II & 0.00       & $\pm$0.09  & $\pm$0.03  & $\mp$0.01 & 0.095 \\ 
&	Ba II& $\pm$0.02  & $\pm$0.05  & $\pm$0.03  & $\mp$0.11 & 0.126 \\ 
\enddata 

\end{deluxetable} 
\end{center}


%% file: tab9.tex

\singlespace
\begin{center}
\begin{deluxetable}{lrlrc}
\tablenum{9}
\tablewidth{0pc}
\tablecolumns{5} 
\tablecaption{Comparison of Abundances for Dwarfs and Giants}
\tablehead{
\colhead{element} & \colhead{Dwarfs} & \colhead{$\sigma$} & \colhead{Giants}
& \colhead{$\sigma$}
}
\startdata
$[$Fe/H$]$        &    +0.12   &    0.04   &   +0.16    &    0.05 \\
$[$O/Fe$]$	  &    $-$0.06 &    0.09   &   $-$0.11  &    0.03 \\ 
$[$Na/Fe$]$	  &    $-$0.01 &    0.03   &   +0.25    &    0.06 \\
$[$Mg/Fe$]$	  &     0.00   &    0.04   &   +0.27    &    0.05 \\
$[$Al/Fe$]$	  &     0.00   &    0.04   &     0.00   &    0.03 \\
$[$Si/Fe$]$	  &    $-$0.01 &    0.03   &   +0.06    &    0.02 \\
$[$Ca/Fe$]$	  &    $-$0.01 &    0.05   &   $-$0.08  &    0.02 \\
$[$Sc/Fe$]$	  &    +0.04   &    0.02   &   $-$0.04  &    0.05 \\
$[$Ti/Fe$]$	  &    $-$0.04 &    0.06   &   $-$0.07  &    0.03 \\
$[$V/Fe $]$	  &    +0.04   &    0.04   &   +0.06    &    0.03 \\
$[$Cr/Fe$]$	  &    +0.00   &    0.03   &   +0.05    &    0.01 \\
$[$Ni/Fe$]$	  &    $-$0.03 &    0.03   &   +0.02    &    0.02 \\
$[$Y/Fe$]$	  &    +0.01   &    0.08   &   $-$0.11  &    0.01 \\
$[$Ba/Fe$]$	  &    +0.11   &    0.04   &   +0.33    &    0.05 \\
\enddata 

\end{deluxetable} 
\end{center}


%% file: line.eqw.tex
\tightenlines

\singlespace
\begin{center}
\begin{deluxetable}{lcccrrr} 
\tablewidth{0pc}
\tablenum{APPENDIX}
\tablecolumns{7} 
\tablecaption{Spectral Lines Used and Measured Equivalent Widths in Three
Stars} 
\tablehead{ 
\colhead{Ion} & \colhead{$\lambda$ (\AA) } & \colhead{Ex.~Pot.~(eV)} &
\colhead{$\log{gf}$} & \colhead{KW 23} & \colhead{KW 208} & \colhead{KW 335} 
} 
\startdata 
{\ion{C}{1}}   &  6587.62  &	 8.53 &	  $-$1.000 & 13.1 & 19.6 & \nodata \\
&   7100.13  &	 8.64 &	  $-$1.020   & 11.8 & \nodata & 17.8 \\	
&   7111.45  &	 8.64 &	  $-$1.080   & 10.8 & 16.6 & 13.5 \\	
&   7113.17  &	 8.65 &	  $-$0.770   & 24.0 & 29.1 & 28.6 \\	
&   7115.17  &	 8.64 &	  $-$0.930   & 25.3 & 32.6 & \nodata \\	
&   7116.96  &	 8.65 &	  $-$0.900   & 18.0 & 27.5 & 19.9 \\	
&   7119.70  &	 8.64 &	  $-$1.220   & 11.3 & \nodata & \nodata \\
\hline
{\ion{O}{1}} & 6158.17  & 10.74 & $-$0.320 & \nodata & 5.7 & \nodata \\
&   7771.94  &    9.11  &   +0.369   & 72.8 & 107.2 & 90.5 \\
&   7774.17  &    9.11  &   +0.223   & 60.2 & 91.7 & 75.1 \\
&   7775.39  &    9.11  &   +0.002   & 53.3 & 78.5 & 61.6 \\
\hline
{\ion{Na}{1}} & 6154.227  & 2.10 & $-$1.660   & 46.6 & 35.5 & 41.6\\
& 6160.751 & 2.10 & $-$1.350   & 65.8 & 51.8 & 64.4 \\
\hline
{\ion{Mg}{1}} & 6965.41  &  5.75 &  $-$1.870   & 27.6 & 21.0 & 26.5 \\
& 7387.70                &  5.75 &  $-1$.200   & 87.4 & 74.8 & 90.2 \\
\hline
{\ion{Al}{1}} & 6698.67  &  3.14 &   $-$1.950   & 25.7 & 18.2 & 22.4 \\
\hline
{\ion{Si}{1}} & 5772.15   &   5.08 &   $-$1.750   & 61.2 & 57.3 & 61.2 \\
&   5948.55 &	  5.08 &   $-$1.225    & 98.2 & 94.1 & 97.2 \\
&   6125.03 &	  5.61 &   $-$1.540    & 39.3 & 37.7 & 39.4 \\
&   6142.49 &	  5.62 &   $-$1.480    & 40.0 & 41.0 & 42.0 \\
&   6155.14 &	  5.62 &   $-$0.840    & 98.4 & 97.1 & 96.2 \\
&   6848.57 &	  5.86 &   $-$1.740    & 20.4 & 19.1 & 20.6 \\
&   7003.57 &	  5.96 &   $-$0.860    & 68.6 & 71.5 & 70.5 \\
&   7005.90 &	  5.98 &   $-$0.680    & 91.4 & 94.6 & 93.8 \\
&   7289.19 &	  5.62 &   $-$0.620    & \nodata & \nodata & \nodata \\
&   7405.79 &	  5.61 &   $-$0.570    & 102.1 & 100.4 & 100.6 \\
&   7760.64 &	  6.20 &   $-$1.280    & 18.3 & 24.9 & 32.6 \\
\hline
{\ion{Ca}{1}} & 6161.30   &  2.52 &  $-$1.270   & 84.8 & 58.9 & 77.8 \\
&   6163.76 &	  2.52 &   $-$1.286   & 81.2 & \nodata & 85.3 \\
&   6166.44 &	  2.52 &   $-$1.140   & 81.9 & 68.7 & 74.6 \\
&   6169.04 &	  2.52 &   $-$0.797   & \nodata & 97.4 & 107.0 \\
&   6169.56 &     2.53 &   $-$0.374   & \nodata & 115.1 & \nodata \\
&   6449.81 &	  2.52 &   $-$0.502   & \nodata & 109.9 & \nodata \\
&   6455.60 &	  2.52 &   $-$1.340   & 69.0 & 51.9 & 67.1 \\
&   6464.68 &	  2.52 &   $-$2.530   & 22.1 & 9.4 & 15.5 \\
&   6471.66 &	  2.53 &   $-$0.638   & \nodata & 96.5 & 102.6 \\
&   6499.65 &	  2.52 &   $-$0.818   & 103.2 & 89.6 & 96.2 \\
\hline
{\ion{Sc}{2}} & 6245.61  &    1.51 &   $-$1.150   & 37.6 & 38.6 & 36.9 \\
&   6604.60   &   1.36   & $-$1.230    & 39.7 & 40.2 & 38.1 \\
\hline
{\ion{Ti}{1}} &   5866.46   &   1.07 &   $-$0.870   & 57.7 & 38.6 & 51.0 \\
&   6126.22 &	  1.07 &   $-$1.460  & 30.5 & 15.5 & 22.6 \\
&   6554.24 &	  1.44 &   $-$1.160  & 23.2 & 7.8  & 13.6 \\
&   6556.08 &	  1.46 &   $-$1.100  & 26.3 & 13.9 & 19.6 \\
&   6599.11 &	  0.90 &   $-$2.060  & 13.5 & \nodata & 9.6 \\
&   6745.55 &	  2.24 &   $-$1.100  & 4.2  & \nodata & \nodata \\
&   7138.93 &	  1.44 &   $-$1.720  & 11.0 & \nodata & 7.1 \\
&   7216.19 &	  1.44 &   $-$1.300  & 26.3 & 10.8 & 15.9 \\
&   7251.72 &     1.43 &   $-$0.860  & 54.7 & 35.4 &  47.7\\ 
&   7440.58 &	  2.25 &   $-$1.080  & 7.6  & \nodata & 9.0 \\
&   7949.15 &	  1.50 &   $-$1.430  & 15.8 & \nodata & 10.5 \\
\hline
{\ion{V}{1}} & 6216.31  &    0.28 &   $-$0.830   & 48.1 & 33.8 & 45.3 \\
&	6504.14   &   1.18  &  $-$0.740   & 18.2 & 11.1 & 15.9 \\
\hline
{\ion{Cr}{1}} &   5783.09  &    3.32  &  -0.500   & 45.4 & 33.1 & 38.0 \\
&   5783.89  &    3.32 &   $-$0.295   & 60.8 & 43.7 & 53.8 \\
&   6330.09  &    0.94 &   $-$2.920   & 39.5 & \nodata & 34.5 \\
&   6501.20  &	  0.98 &   $-$3.660   & 12.6 & \nodata & 12.0 \\	 
&   6605.57  &    4.14 &   $-$0.810   & \nodata & \nodata & \nodata \\
&   6630.03  &    1.03 &   $-$3.560   & 12.3 & 9.3 & 10.7 \\
&   6978.48  &	  3.46 &    +0.143    & 81.1 & 65.7 & 70.9 \\	 
&   6979.81  &	  3.46 &   $-$0.460   & 49.2 & 38.5 & 43.3 \\	 
&   6980.91  &	  3.46 &   $-$1.120   & 16.4 & 10.3 & 9.6 \\
\hline
{\ion{Fe}{1}} & 5775.08 &     4.22 &   $-$1.300 & 70.5 & 59.5 & 65.4  \\
&   5809.22 &	  3.88 &   $-$1.760  & 64.5 & 51.6 & 59.0 \\
&   5849.69 &	  3.69 &   $-$2.970  & 11.2 & 5.1 & 8.8 \\
&   5852.22 &	  4.55 &   $-$1.220  & 51.9 & 41.4 & 47.7 \\
&   5855.08 &	  4.61 &   $-$1.560  & 28.5 & 22.3 & 25.5 \\
&   5856.09 &	  4.29 &   $-$1.600  & 42.5 & 33.4 & 39.1 \\
&   5858.78 &	  4.22 &   $-$2.190  & 17.6 & 14.3 & 17.6 \\
&   5859.59 &	  4.55 &   $-$0.610  & 83.3 & 76.2 & 82.8 \\
&   5861.11 &	  4.28 &   $-$2.340  & 10.4 & 7.8  & 8.9   \\
&   5862.36 &	  4.55 &   $-$0.420  & 101.8 & 93.3 & 102.1 \\
&   6027.05 &	  4.08 &   $-$1.150  & 73.8 & 66.6 & 72.8 \\
&   6055.99 &	  4.73 &   $-$0.460  & 84.8 & 78.9 & 82.2 \\
&   6127.90 &	  4.14 &   $-$1.399  & 58.7 & 50.7 & 55.2 \\
&   6151.62 &	  2.18 &   $-$3.299  & 59.8 & 44.1 & 55.8 \\
&   6157.73 &	  4.07 &   $-$1.270  & 71.9 & 65.0 & 66.7 \\
&   6159.38 &	  4.61 &   $-$1.880  & 16.8 & 10.9 & 13.8 \\
&   6165.36 &	  4.14 &   $-$1.580  & 53.3 & 42.6 & 51.1 \\
&   6173.34 &	  2.22 &   $-$2.880  & 81.1 & 69.5 & 75.8 \\
&   6180.20 &	  2.73 &   $-$2.620  & 68.7 & 58.2 & 64.9 \\
&   6213.44 &	  2.22 &   $-$2.660  & 98.8 & 81.0 & 96.8 \\
&   6219.29 &	  2.20 &   $-$2.433  & 107.4 &93.9  & 103.2 \\
&   6226.74 &	  3.88 &   $-$2.220  & 37.1 & 27.2 & 36.8 \\
&   6229.23 &	  2.84 &   $-$3.020  & 50.5 & 37.2 & 47.6 \\
&   6240.65 &	  2.22 &   $-$3.200  & 59.8 & 45.6 & 55.8 \\
&   6270.23 &	  2.86 &   $-$2.690  & 59.8 & 54.6 & 57.4 \\
&   6271.28 &	  3.33 &   $-$2.840  & 35.0 & 24.6 & 31.2 \\
&   6290.97 &	  4.73 &   $-$0.760  & 80.8 & 67.9 & 73.3 \\
&   6322.68 &	  2.59 &   $-$2.450  & 88.9 & 73.3 & 81.4 \\
&   6330.85 &	  4.73 &   $-$1.280  & 44.4 & 37.7 & 39.0 \\
&   6335.34 &	  2.20 &   $-$2.230  & \nodata & 98.6 & 106.9 \\
&   6344.15 &	  2.43 &   $-$2.900  & 68.8 & 60.7 & 69.0 \\
&   6380.75 &	  4.19 &   $-$1.410  & 62.2 & 49.5 & 58.9 \\
&   6392.54 &	  2.28 &   $-$4.030  & 26.2 & 13.4 & 16.9 \\
&   6481.87 &	  2.28 &   $-$2.980  & 77.1 & 61.8 & 67.4 \\
&   6498.94 &	  0.96 &   $-$4.690  & 56.5 & 38.3 & 49.3 \\
&   6581.22 &	  1.48 &   $-$4.790  & 28.2 & 19.1 & 23.9 \\
&   6591.33 &	  4.59 &   $-$2.070  & 12.1 & 7.9  & 10.8 \\
&   6608.04 &	  2.28 &   $-$4.020  & 28.1 & 12.9 & 17.7 \\
&   6625.04 &	  1.01 &   $-$5.350  & 27.9 & 11.0 & 17.9 \\
&   6627.56 &	  4.55 &   $-$1.610  & 40.1 & 28.6 & 32.1 \\
&   6703.57 &	  2.76 &   $-$3.130  & 45.6 & 34.0 & 39.0 \\
&   6705.12 &	  4.61 &   $-$1.170  & 57.8 & 49.7 & 51.1 \\
&   6726.67 &	  4.61 &   $-$1.160  & 57.0 & 49.2 & 55.5 \\
&   6750.15 &	  2.42 &   $-$2.620  & 85.8 & 70.1 & 80.4 \\
&   7071.87 &	  4.61 &   $-$1.700  & 32.5 & 29.6 & 32.4 \\
&   7107.47 &	  4.19 &   $-$2.070  & 30.1 & 20.1 & 27.4 \\
&   7127.57 &	  4.99 &   $-$1.250  & 36.0 & 28.4 & 34.7 \\
&   7130.93 &	  4.22 &   $-$0.740  & 108.1 &93.5 & 99.8 \\
&   7132.98 &	  4.07 &   $-$1.770  & 51.8 & 44.3 & 50.4 \\
&   7142.52 &	  4.95 &   $-$1.090  & 47.6 & 38.9 & 44.3 \\
&   7155.63 &	  5.01 &   $-$1.090  & 48.1 & 36.9 & 43.6 \\
&   7745.52 &	  5.08 &   $-$1.180  & 31.0 & 32.3 & 34.9 \\
&   7746.60 &	  5.06 &   $-$1.290  & 28.8 & 28.3 & 26.7 \\
&   7751.11 &	  4.99 &   $-$0.770  & 58.2 & 57.8 & 57.5 \\
&   7879.78 &	  5.03 &   $-$1.650  & 15.3 & 9.3  & 11.3 \\
\hline
{\ion{Fe}{2}} & 6149.25 &     3.89 &   $-$2.724   & 39.6 & 48.3 & 43.5 \\
&   6247.56 &	  3.89 &   $-$2.329    & 58.4 & 71.9 &  64.9 \\   
&   6456.39 &	  3.90 &   $-$2.075    & 70.4 & 82.0 &  77.2 \\   
&   6516.08 &	  2.89 &   $-$3.380    & 60.3 & 65.3 &  62.7 \\   
&   7224.46 &	  3.89 &   $-$3.243    & 18.8 & 25.5 &  21.5 \\   
&   7711.73 &	  3.90 &   $-$2.450    & 44.6 & 57.7 &  51.6 \\
\hline
{\ion{Ni}{1}} &   5847.00  &    1.68 &   $-$3.430   & 29.5 & 18.6 & 23.3 \\
&   5857.75 &	  4.17 &   $-$0.390  & 67.2 & 67.5 & 73.8 \\
&   6108.11 &	  1.68 &   $-$2.450  & 73.6 & 62.1 & 68.4 \\
&   6130.14 &	  4.09 &   $-$1.040  & 27.0 & 19.8 & 24.5 \\
&   6133.98 &	  4.09 &   $-$1.770  & 9.8  & 5.6  & 4.5 \\
&   6175.37 &	  4.09 &   $-$0.590  & 60.4 & 51.8 & 59.4 \\
&   6176.82 &	  4.09 &   $-$0.370  & 74.4 & 62.2 & 70.2 \\
&   6177.25 &	  1.83 &   $-$3.530  & 21.3 & \nodata & 17.9 \\
&   6204.61 &	  4.09 &   $-$1.170  & 31.3 & 23.4 & 23.8 \\
&   6482.80 &	  1.94 &   $-$2.630  & 50.2 & \nodata & \nodata \\
&   6586.31 &	  1.95 &   $-$2.810  & 51.9 & 37.5 & 46.6 \\
&   6598.61 &	  4.23 &   $-$1.020  & 30.6 & 24.9 & 31.8 \\
&   6842.04 &	  3.66 &   $-$1.520  & 32.2 & 29.0 & 30.6 \\
&   7001.55 &	  1.93 &   $-$3.620  & 15.2 & 10.3 & 11.4 \\
&   7122.21 &	  3.54 &   $-$0.050  & 122.4 & 105.6 & 115.0 \\
&   7385.24 &	  2.74 &   $-$2.070  & 54.4  & 46.4  & 52.4 \\
&   7393.61 &	  3.61 &   $-$0.040  & 108.3 & 100.9 & 107.3 \\
&   7414.51 &	  1.99 &   $-$2.440  & 77.2  & 57.8  & 70.0 \\
&   7422.29 &     3.63 &   $-$0.010  & 109.4 & 100.7 & 104.9 \\
&   7525.12 &	  3.63 &   $-$0.690  & 81.2  & 71.5  & 79.1 \\
&   7555.61 &	  3.85 &    +0.060   & 103.0 & 93.7  & 102.7 \\
&   7574.05 &	  3.83 &   $-$0.630  & 75.3  & 65.5  & 69.0 \\
&   7714.31 &	  1.93 &   $-$1.800  & \nodata & 98.8 & 113.8 \\
&   7715.59 &	  3.70 &   $-$1.140  & 61.9  & 55.7  & \nodata \\
&   7727.62 &	  3.68 &   $-$0.150  & 106.5 & 90.2  & 100.0 \\
&   7748.89 &	  3.70 &   $-$0.180  & 101.8 & 93.1  & 95.8 \\  
\hline
{\ion{Y}{1}} &   6435.050  &   0.57 & $-$0.830   & 3.8 & \nodata & \nodata \\
&   6687.51  &    0.50 &   $-$0.430   & 11.5 & 3.0 & 6.1 \\
\hline
{\ion{Y}{2}} &   6613.73    &  1.75  & $-$1.110   & 17.6 & 11.8 & 16.4 \\
&   7264.16  &    1.84  &  $-$1.500   & 10.6 & 4.2 & 6.4 \\ 
&   7881.88  &    1.84  &  $-$0.570   & 40.0 & 40.0 & 36.0 \\ 
\hline
{\ion{Ba}{2}} & 5853.70   &   0.60  &  $-$0.970   & 72.6 & 74.0  & 75.3 \\
\hline
\enddata

\end{deluxetable} 
\end{center}
